\documentclass[12pt,linenumbers]{spieman}  
\usepackage{amsmath,amsfonts,amssymb}
\usepackage{graphicx}
\usepackage{setspace}
\usepackage{tocloft}
\usepackage[left]{lineno}
\def\escape~{\textit{ESCAPE}}

\title{The Extreme-ultraviolet Stellar Characterization for Atmospheric Physics and Evolution (ESCAPE) Mission: Motivation and Overview}

\author[a]{Kevin France}
\author[a]{Brian Fleming}
\author[a]{Allison Youngblood}
\author[a]{James Mason}
\author[b]{Jeremy J. Drake}
\affil[a]{Laboratory for Atmospheric and Space Physics, University of Colorado Boulder, Boulder, CO 80303}
\affil[b]{Smithsonian Astrophysical Observatory, 60 Garden Street, Cambridge MA 02138}
\author[c]{Ute V. Amerstorfer}
\affil[c]{Space Research Institute, Austrian Academy of Sciences, Schmiedlstrasse 6, 8042 Graz, Austria}

\author[d]{Martin Barstow}
\affil[d]{School of Physics \& Astronomy, University of Leicester, University Road, Leicester LE1 7RH, UK}

\author[e]{Vincent Bourrier}
\affil[e]{Observatoire Astronomique de l'Université de Genève, Chemin Pegasi 51b, CH-1290 Versoix, Switzerland}

\author[f]{Patrick Champey}
\affil[f]{NASA/Marshall Space Flight Center, Huntsville, Alabama 35812, USA}

\author[c]{Luca Fossati}

\author[g]{Cynthia S. Froning}
\affil[g]{McDonald Observatory, University of Texas at Austin, Austin, TX 78712, USA}

\author[h]{James C. Green}
\affil[h]{Center for Astrophysics and Space Astronomy, University of Colorado Boulder, Boulder CO 80303, USA}

\author[i]{Fabien Gris\'e}
\affil[i]{Dept. of Astronomy \& Astrophysics, Pennsylvania State University, University Park, PA 16802, USA}

\author[j]{Guillaume Gronoff}
\affil[j]{Science Systems and Application Inc. / Nasa Langley Research Center, Hampton VA 23681, USA}

\author[a]{Timothy Hellickson}

\author[k]{Meng Jin}
\affil[k]{Lockheed Martin Solar \& Astrophysics Laboratory (LMSAL), Palo Alto, CA 94304, USA / SETI Institute, Mountain View, CA 94043, USA}

\author[l]{Tommi T. Koskinen}
\affil[l]{Lunar and Planetary Laboratory, University of Arizona, Tucson, AZ 85721--0092}

\author[m]{Adam F. Kowalski} 
\affil[m]{National Solar Observatory, University of Colorado Boulder, 3665 Discovery Drive, Boulder, CO 80303, USA / Department of Astrophysical and Planetary Sciences, University of Colorado, Boulder, 2000 Colorado Ave, CO 80305, USA / Laboratory for Atmospheric and Space Physics, University of Colorado Boulder, 3665 Discovery Drive, Boulder, CO 80303, USA.}

\author[a]{Nicholas Kruczek}

\author[n]{Jeffrey L. Linsky}
\affil[n]{JILA, University of Colorado and NIST, UCB 440, Boulder CO, 80309, USA}

\author[o]{Sarah J. Lipscy}
\affil[o]{Ball Aerospace, Boulder CO 80301, USA}

\author[i]{Randall L. McEntaffer}

\author[f]{David E. McKenzie}

\author[i]{Drew M. Miles}

\author[a]{Tom Patton}

\author[f]{Sabrina Savage}

\author[p]{Oswald Siegmund}
\affil[p]{Space Sciences Laboratory, University of California, Berkeley CA 94720, USA}

\author[a]{Constance Spittler}

\author[a]{Bryce W. Unruh}

\author[a]{M\'{a}ire Volz}

\cftpagenumbersoff{figure}
\cftpagenumbersoff{table} 
\begin{document} 
\maketitle

\begin{abstract} 

The {\it Extreme-ultraviolet Stellar Characterization for Atmospheric Physics and Evolution} (\escape~) mission is an astrophysics Small Explorer employing ultraviolet spectroscopy (EUV: 80 – 825 \AA\ and FUV: 1280 – 1650 \AA) to explore the high-energy radiation environment in the habitable zones around nearby stars.  \escape~ provides the first comprehensive study of the stellar EUV and coronal mass ejection environments which directly impact the habitability of rocky exoplanets.  In a 20 month science mission, \escape~ will provide the essential stellar characterization to identify exoplanetary systems most conducive to habitability and provide a roadmap for NASA’s future life-finder missions.  \escape~ accomplishes this goal with roughly two-order-of-magnitude gains in EUV efficiency over previous missions.  \escape~ employs a grazing incidence telescope that feeds an EUV and FUV spectrograph. The \escape~ science instrument builds on previous ultraviolet and X-ray instrumentation, grazing incidence optical systems, and photon-counting ultraviolet detectors used on NASA  astrophysics, heliophysics, and planetary science missions. The \escape~ spacecraft bus is the versatile and high-heritage Ball Aerospace BCP-Small spacecraft.  Data archives will be housed at the Mikulski Archive for Space Telescopes (MAST).  

\end{abstract}


{\noindent \footnotesize\textbf{*}Kevin France,  \linkable{kevin.france@colorado.edu} }

\begin{spacing}{2}   

\section{Introduction}
\label{sec:intro}

Owing to their large number and strong atmospheric impacts, EUV photons are the primary agent driving atmospheric escape on planets orbiting cool stars (F, G, K, and M stars; approximately 2 – 0.1 solar masses). The stability of Earth-like atmospheres depends critically on the EUV irradiance~\cite{johnstone15,johnstone21}.  Higher EUV flux from the young Sun~\cite{tu15} could have led to 10$\times$ greater oxygen loss rates and 90$\times$ greater carbon loss rates on the early Martian atmosphere, compared to present day levels, by increasing the suprathermal or “hot” population of these atoms~\cite{amerstorfer17}.  On highly irradiated planets, the outflow is sufficiently rapid that heavier atmospheric species (e.g., N, C, O, Mg, Fe) can be dragged along through collisions with the lighter hydrogen, as observed even on hot Jupiters~\cite{vidal04,linsky10,ballester15,fossati10,haswell12,sing19,cubillos20}.  Free electrons produced by stellar EUV photons can also attain altitudes much greater than ions, producing an ambipolar electric field that leads to a non-thermal ionospheric outflow in O$^{+}$ and N$^{+}$ winds~\cite{kulikov06,lichtnegger16,dong17,airapetian20}. 
Consequently, a major open question as astronomers embark on the spectral characterization of rocky exoplanets in the  coming decade with $JWST$ is whether terrestrial atmospheres can survive the radiation environments around other stars \cite{gronoff2020}.

Previous EUV observatories have lacked the sensitivity to survey exoplanet host stars in this spectral band, and as a result, widely discrepant techniques have been employed to approximate the critical stellar inputs into  exoplanet atmosphere models. In this paper,  we present the motivation for and the mission overview of the {\it Extreme-ultraviolet Stellar Characterization for Atmospheric Physics and Evolution} (\escape~) mission.  \escape~ explores the habitable zone (HZ) radiation environment around more than 200 nearby stars, filling the observational gap in planetary atmosphere evolution models, models that enable us to predict which star-planet systems are most promising for developing and maintaining habitable conditions.  These nearby stellar systems make up the target list for future NASA missions focused on spectroscopic characterization of rocky planets. 

\escape~ investigates stars of spectral types F, G, K, and M to be responsive to all terrestrial exoplanet characterization missions currently in implementation or formulation.  \textcolor{black}{Terrestrial exoplanet research with }$JWST$ in the 2020s and the Origins Space Telescope (OST) \textcolor{black}{or an IR probe} in the 2030s \textcolor{black}{are best suited to study} transiting exoplanets around M dwarfs.  Direct detection missions of the 2030s and 2040s such as Habex, LUVOIR, \textcolor{black}{and the 6-meter UVOIR mission recently recommended by the Astro2020 Decadal Survey} will be optimized for exo-Earths orbiting more massive F, G, and K stars.   \escape~’s observing program provides the stellar EUV context for whatever technical and programmatic approach the community adopts to identify and characterize habitable worlds.  The \escape~ data set will allow the astronomical community to predict the most promising star-planet systems for life detection missions in the coming decades, establishing a roadmap for future observing resources to explore habitable worlds.



%
\section{SCIENCE OBJECTIVES AND IMPLEMENTATION}
\label{sec:obs}

\escape~’s primary science goal is to identify and understand those star-planet systems that are conducive to the formation of habitable environments.  The \escape~ mission addresses this goal by answering three key science questions: 
\begin{enumerate}
    \item What is the EUV irradiance in the habitable zone?  
    \item How does stellar EUV irradiance evolve in time?   
    \item What are the properties of stellar coronal mass ejections? 
\end{enumerate}

\escape~ addresses these science questions through a comprehensive observing program supported by state-of-the-art planetary atmosphere models that take \escape~'s observables as inputs, and are used to estimate atmospheric mass loss rates from representative terrestrial planets.  The \escape~ instrument: (1) achieves $\approx$~25--100$\times$ the EUV efficiency of previous missions, resolving order-of-magnitude uncertainties on exoplanet radiation environments, (2) explores EUV variability from flares, stellar rotation, and stellar evolution with its photon-counting detector, monitoring survey, and sample of F, G, K, and M stars with a range of ages, and (3) executes the first survey of stellar coronal mass ejections (CMEs) using techniques validated in Sun-as-a-star measurements by NASA’s $SDO$-EVE instrument, providing direct constraints on exoplanet particle environments.

\section{High-Energy Radiation Environments: EUV Photons,  Variability, and CMEs}
\label{sec:obs}

\subsection{Stellar EUV Irradiance}

Optical and near-infrared photons heat the surface and troposphere. NUV (1800 – 3200 \AA), FUV (912 – 1800 \AA), and X-ray (5 – 100 \AA) photons are absorbed in the middle and upper atmosphere where they photo-dissociate molecules and ionize heavy elements.  EUV photons (100–911 \AA) are absorbed high in the atmosphere (i.e., in the thermosphere) where they ionize atoms and molecules. Liberated electrons collisionally heat the surrounding gas, increasing the scale height of the atmosphere, driving ambipolar ion mass loss, and potentially leading to the formation of a hydrodynamic outflow.  Rapid atmospheric loss can lead to both desiccation and the build-up of abiotic O$_{2}$ atmospheres that complicate biosignature searches~\cite{ramirez14,luger15,wordsworth18}.

\begin{figure*}[htpb]
   \centering
   \includegraphics[scale=.50,clip,trim=0mm 0mm 0mm 0mm,angle=0]{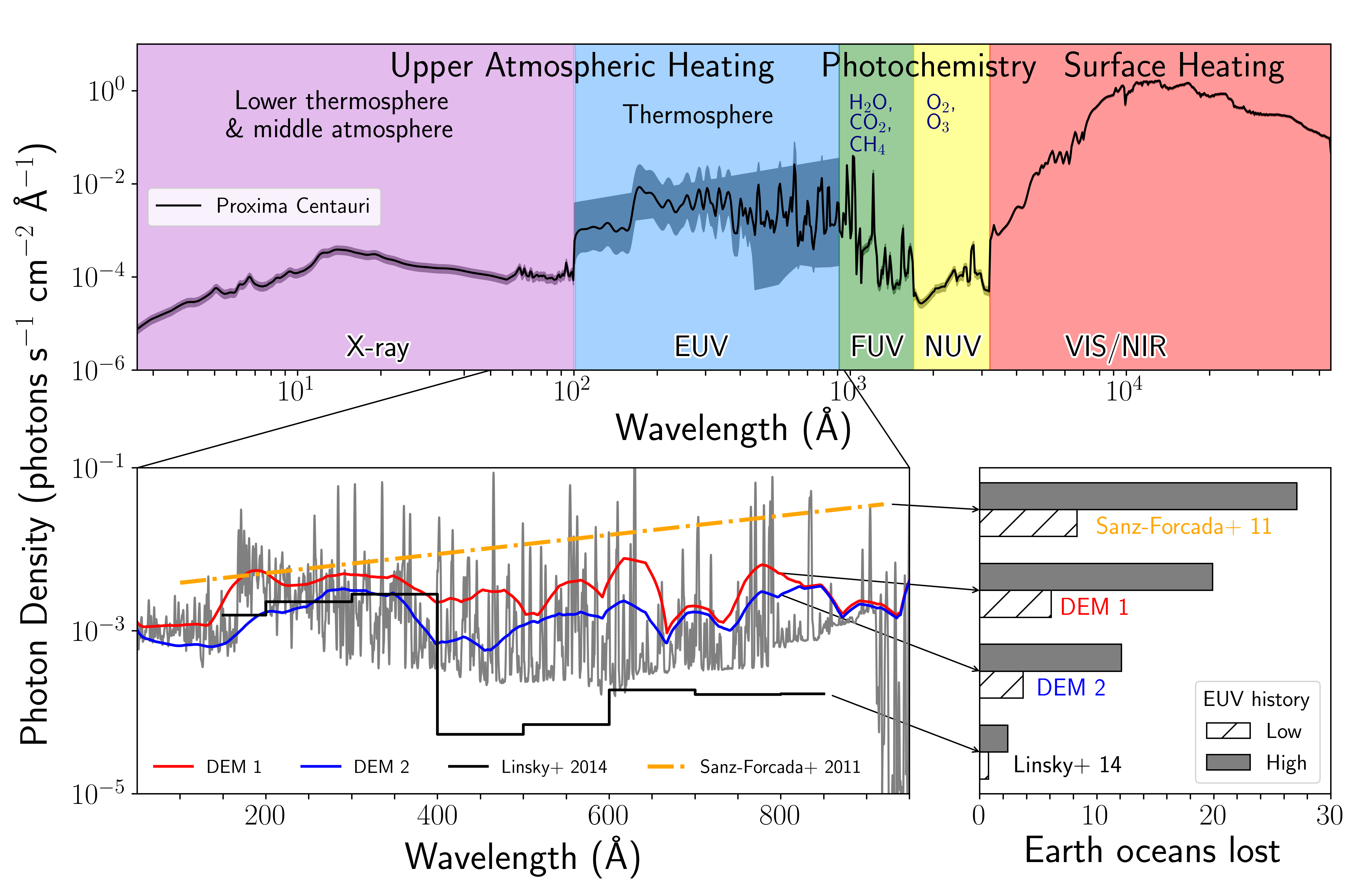}
   \caption{(top) Composite spectrum of the nearby planet-hosting M dwarf Proxima Cen,  adapted from France et al. (2016) and Loyd et al. (2016)\nocite{france16,drake20,linsky14,forcada11,ribas17,loyd16}.  The width of the shaded region indicates the approximate uncertainty on the intrinsic flux level.     (lower left) Different reconstructions of the EUV spectrum of Proxima Cen show factors of 3 -- 100 flux discrepancies:  differential emission measure models in red, blue, and gray (Drake et al. 2020); scaling relations based on only FUV data in black (Linsky et al. 2014) and only X-ray data in orange (Sanz-Forcada et al. 2011).   The corresponding atmospheric hydrogen mass lost from an Earth-like planet (in units of Earth oceans; 1 Earth ocean~=~1.4~$\times$~10$^{24}$ g) from 10 Myr to 4.8 Gyr is shown at lower right for high (shaded gray) and low (hatched) EUV histories of typical M dwarfs (Ribas et al. 2017).  
   }
\end{figure*}

EUV photons are the key drivers for atmospheric mass-loss for three primary reasons:  (1) EUV photons are absorbed in the highest (lowest density) layers of the atmosphere, where radiative losses (proportional to density squared) are minor and the heating efficiency is highest, (2) the ionization  cross-sections for dominant upper atmospheric atomic species (e.g., H, N, O) peak at EUV wavelengths and (3) there are many more EUV photons than X-rays available on all types of stars to drive this heating.  In the quiet Sun, the EUV/X-ray photon production ratio is $\sim$90~\cite{woods09}.  For optically inactive early M dwarfs the EUV/X-ray photon ratio is $\sim$40~\cite{fontenla16}, and even for an active late M dwarf like Proxima Cen, the EUV/X-ray photon ratio is $\sim$16 (Figure 1).

 The discovery of rocky planets in the HZs of nearby stars, e.g., Proxima Cen b and the TRAPPIST-1 planets, has motivated new atmospheric mass loss calculations that highlight the need for improved EUV irradiance data to estimate their long-term habitability.  Garcia-Sage et al. (2017)\cite{garcia17} and Airapetian et al. (2017)\cite{airapetian17} presented studies of EUV-driven ion loss from Earth-like planets orbiting M dwarfs.  They found mass loss rates  several orders of magnitude higher than that of present-day Earth for EUV fluxes 10 - 20 times the current day EUV solar irradiance. Their models indicate that elevated EUV fluxes, 
augmented by persistent flares or the long pre-main-sequence phase of M dwarfs, could effectively render some Earth-like planets around M dwarfs devoid of an atmosphere, in the absence of internal or external resupply of volatiles. A factor of ten in the uncertainty of incident EUV flux can lead to three orders of magnitude in the uncertainties of ion escape rates~\cite{garcia17}, fundamentally shifting our predictions for which exoplanets are the best candidates for long-lived habitable atmospheres. 

Figure 1 shows the results of analogous thermal atmospheric escape calculations for Proxima Cen b under different plausible EUV radiation strengths and histories.  Model calculations yield a factor of ~$\sim$~30 spread in the total atmospheric mass loss from the planet over time, driven almost entirely by \textcolor{black}{what assumptions are used to estimate} the current EUV flux and its evolution earlier in the star’s history.

Despite their importance, the EUV luminosity, flare rate, and evolution of other stars are almost completely unconstrained and can only be empirically determined with direct measurements of the EUV flux and temporal variability of a sample of stars with a range of ages and activity levels.  Existing EUV spectra of exoplanet host stars are scarce. The only previous EUV astronomy mission, the {\it Extreme Ultraviolet Explorer} ($EUVE$; Bowyer \& Malina. 1991),\nocite{bowyer91} obtained spectra of $\sim$15 cool main sequence stars, including 5 M dwarfs.  $EUVE$'s observations were heavily biased toward the most active stars. The very modest $\leq$ 2 cm$^{2}$ peak effective area of the $EUVE$ spectrometers precluded useful spectroscopic observations of stars with more solar-like activity, except for the nearby $\alpha$ Cen system and the F4 subgiant Procyon~\cite{drake95,drake97}. No observed EUV spectra exist of the optically inactive M dwarfs (\textcolor{black}{defined here as} Ca II H \& K equivalent widths $<$ 1.0 \AA)\cite{walkowicz09,france16}, that will be optimal for atmospheric studies with $JWST$, 30-m telescopes, and the Origins Space Telescope.  The lack of direct EUV data hampers our ability to understand how habitable atmospheres evolve with time, and to design truly definitive biosignature searches.\nocite{walkowicz09,france16}


\subsection{EUV Evolution and Flares }

Solar EUV emission varies by factors of up to $\sim$100 on minute timescales due to flares and by factors of $\sim$6 on year timescales due to the solar cycle~\cite{woods12}. Different emission lines and continua vary by different factors on all timescales, while connections between the flare energy and frequency distribution of white light flares (such as those observed with $Kepler/K2$ and $TESS$) and UV flares remain poorly constrained by observation and theory~\cite{segura10,davenport16,guenther20,howard20}.

Estimates of the EUV response of the Great AD Leo Flare of 1985~\cite{hawley91} observed with the {\it International Ultraviolet Explorer}, $IUE$, indicate that the atmospheres of planets orbiting very active stars with frequent flares never achieve a steady state because the timescales for atmospheric recovery are much longer than the time between successive flares~\cite{venot16}. $EUVE$ provided constraints on flare frequencies for a small sample of young, active stars~\cite{audard00}, however, the lack of direct constraints on EUV flare activity on older stars (e.g., France et al. 2020) hinders our ability to model the stability of potentially habitable atmospheres over timescales required for the initial development of surface life ($\sim$~1.7 Gyr; Jones \& Sleep 2010).\nocite{jones10,france20} \escape~ employs a photon-counting detector to record all observations in a `time-tagging' mode that provides temporal resolution ($<$ 20s, limited by photon statistics, the \textcolor{black}{instrumental timing resolution is $<$ 0.1s}) higher than the characteristic UV flare duration on low-mass stars ($\sim$300s; Hawley et al. 2003; Loyd \& France 2014; Loyd et al. 2018a).\nocite{hawley03,loyd14,loyd18a}

Characterization of the EUV flare frequency and spectral variability on stellar evolutionary timescales requires (1) directly collecting EUV spectra over long enough temporal baselines to monitor variability and (2) a broad sample of stars with a range of masses and ages (activity levels, e.g., West et al. 2015).\nocite{west15}   The required monitoring observations cannot be collected from the ground because of the lack of optical features whose behavior is closely linked to the EUV emission. For example, ``inactive" M dwarfs that show little optical lightcurve variation at the $<$1\% level can display factors of~$\sim$~50 increases in the FUV and X-ray~\cite{france16, loyd18a}, and we expect the behavior to be similar at EUV wavelengths.  \escape~ employs a dedicated monitoring program to sample the flare frequency and energy distribution for individual systems across spectral classes and ages (Section 5).

\subsection{Detecting Stellar CMEs}

High-energy particles have a major influence on the evolution of exoplanetary atmospheres, both through the actions of stellar winds and impulsive events (e.g., CMEs).  While stellar winds play a part in atmospheric escape, CMEs are thought to be the dominant source of long-term instability in planetary atmospheres~\cite{khodachenko07,lammer07,cherenkov17}. 
The influence of CMEs depends strongly on their frequency and ability to break out of coronal magnetic confinement; \escape~ will measure the frequencies of CMEs on solar-type stars and search for CME breakout from M dwarfs.  

\begin{figure*}
\centering 
\includegraphics[width=0.82\textwidth]{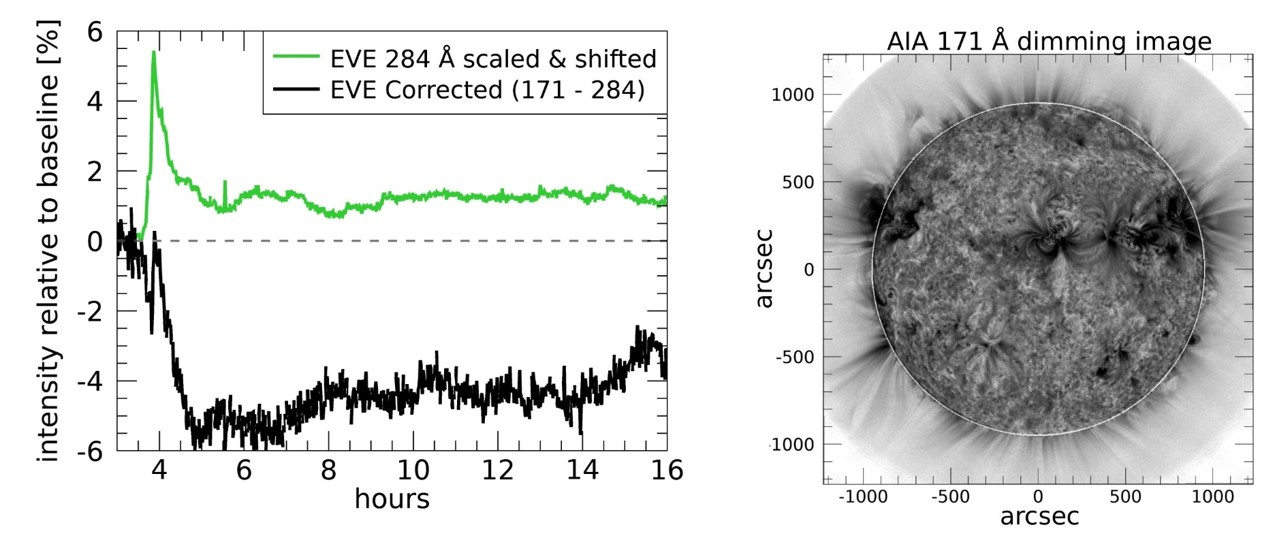}
    \caption{Observation of coronal dimming in full-disk solar spectra (left) and angularly resolved filter imaging (right).  The spectrally-resolved lightcurves from $SDO$-EVE show flaring lines (Fe XV 284 \AA, green) and the corrected Fe IX 171 \AA\ dimming profile in black (Mason et al. 2016). \escape~ is sensitive to similar events on F, G, and K dwarfs to a distance of 6 pc.  More than 90 candidate events are expected in the DEEP survey (Section 5.2).\nocite{mason14,mason16,mason19}
    }
    \label{OvsB}
\end{figure*}

While the main source of heat in the upper atmosphere of planets is the EUV flux, Joule heating and particle precipitation heating could dominate the chemical impact of impulsive events (e.g., Segura et al. 2010; Tilley et al. 2019).\nocite{segura10,tilley19}  CMEs deliver heat to the upper atmospheres of orbiting planets through charge exchange reactions~\cite{chassefiere96}. Pickup and sputtering processes can also be major sources of atmospheric loss from direct particle precipitation and stellar wind enhancements due to CMEs.   CMEs contribute to atmospheric mass loss on solar system planets~\cite{jakosky15,jakosky18} and to space weather, leading several authors to propose an expanded definition of the habitable zone that takes into account space weather impacts (e.g., Airapetian et al.  2017, 2020) as more meaningful than the traditional liquid water HZ. 

Approximately 90\% of large (X-class) solar flares are associated with CME-like particle eruptions~\cite{yoshiro06}, however, this connection has not been borne out with recent observations of stellar flares~\cite{osten15,odert17,crosley18}.
While the EUV irradiance is increased by a large flare rate, CMEs and accelerated particles may have much greater impacts on atmospheric photochemistry and stripping than the flare radiation itself~\cite{segura10,lammer07}.

For the Sun, we are able to observe CMEs directly with coronagraphs and in situ measurements. These traditional methods are not feasible for stellar CMEs in the near-term (although see Haisch et al. 1983 and Moschou et al. 2017).\nocite{haisch83,moschou17} However, other techniques for detecting solar CMEs have been developed. \textcolor{black}{Harra et al. (2016)}\cite{harra16} reviewed these techniques and determined that coronal dimming, described below, is the only feature consistently associated with CMEs that can be employed to detect and quantify these events on other stars.

Solar coronal dimming studies have primarily relied upon spatially resolved EUV images to characterize the transient voids left behind in the corona as a CME departs, giving rise to the observed flux dimming (e.g., Sterling and Hudson 1997, Aschwanden et al. 2009, Reinard and Biesecker 2009; Dissauer et al. 2018a,b). The {\it Solar Dynamics Observatory}-EVE instrument ($SDO$-EVE; Woods et al. 2012) demonstrated that dimming can also be characterized in disk-integrated EUV spectra~\cite{woods11,mason14,mason16,mason19}.  Similar dimming events have recently been reported in X-ray lightcurves of active stars and an archival $EUVE$ spectrum of the young K star AB Dor~\cite{veronig21}.  \escape~ employs the same `Sun-as-a-star' stellar CME characterization techniques validated on $SDO$-EVE (Figure 2) to measure the frequency of CMEs on nearby stars for the first time.\nocite{sterling97,dissauer18a,dissauer18b,aschwanden09,reinard09}

The dimming light curve also contains information about the kinematics of the CME that produced it. Solar dimming lightcurves have been calibrated by in-situ proton measurements and coronagraphic imaging of CMEs.  The depth of the dimming event is directly related to the mass of the ejected CME (the dispersal of the quiescent corona greatly reduces the emission measure, which scales as density squared), whereas the slope of the dimming lightcurve indicates the CME propagation velocity~\cite{mason16,mason19}.  Therefore, observations of coronal dimming events on solar-type stars can provide direct constraints on the physical properties of the CME. 

\begin{figure}[htbp]
   \centering
   \includegraphics[scale=.45,clip,trim=0mm 0mm 0mm 0mm,angle=0]{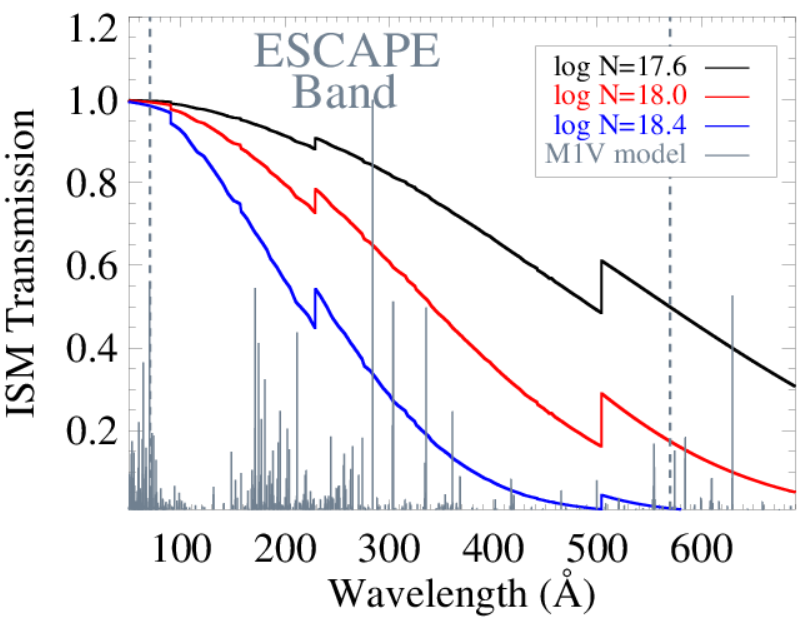}

   \caption{Transmission of the local interstellar medium for H I column densities typical of stars inside 30 pc (10$^{17.6-18.4}$ cm$^{-2}$; Wood et al. 2005). These curves demonstrate that the ISM is more than 20\% transparent to EUV photons over most of the \escape~ \textcolor{black}{G20 and G40 bandpasses} for most stars inside 30 pc. }

\end{figure}

\section{Interstellar Attenuation and Flux Reconstruction}

Dust and molecular hydrogen extinction is negligible within the Local Cavity (Lehner et al. 2003; also historically referred to as the Local Bubble), meaning that the only significant sources of line-of-sight \textcolor{black}{opacity (or, “reddening”)} are neutral and low-ionization atomic gases~\cite{redfield08}.\nocite{lehner03}  The \escape~ FUV bandpass covers 1280 – 1650 Å (excluding the strong Ly$\alpha$ line); the only significant sources of interstellar opacity in the FUV band are neutral oxygen and singly ionized carbon and silicon, which impact measurements of low-ionization lines in \escape~’s FUV range (e.g., C II).  In \escape~’s EUV range, neutral hydrogen (H I), neutral helium (He I) and ionized helium (He II) contribute, with the H I column density being the controlling parameter for the total line-of-sight EUV extinction (Figure 3).  Despite the physical complexity of the local ISM, integrated H I column density rarely exceeds 10$^{18.4}$ cm$^{-2}$ within 30 pc of the Sun~\cite{wood05} and do not reach 10$^{19}$ cm$^{-2}$ routinely until $d$~$>$~80 pc, beyond \escape~'s baseline mission target distance.

\textcolor{black}{ESCAPE employs three techniques to measure or estimate the interstellar hydrogen column density (N(HI)) towards its targets and therefore to correct for the ISM attenuation in its spectra.   First, high-fidelity direct measurements of N(HI) are routinely acquired with the $HST$-STIS instrument (or archival data from $HST$-GHRS).  Direct Lyman-$\alpha$ measurements of N(HI) yield uncertainties of $<$ 0.1 dex (see, e.g., Table 2 of Wood et al.  2005; Tables 2 and 3 of Youngblood et al. 2016).\nocite{youngblood16,loyd16,france16b}  Second, an all-sky n(HI) map created from ensemble archival N(HI) measurements has been created and subsequently validated by comparing new observations from HST~\cite{wood21} against the predicted N(HI) from the all-sky map, finding agreement between direct observation and all-sky model prediction of $\leq$~0.25 dex.   This level of interstellar column uncertainty yields absolute flux uncertainties $\leq$~35\% for wavelengths shorter than 300~\AA\ and column densities typical of the local ISM within 30 pc ($\leq$~10$^{18.4}$ cm$^{-2}$).  Finally, these empirical techniques are augmented by three-dimensional ISM models based on empirical studies of the local ISM and publicly available to the community (Redfield \& Linsky 2000, 2008).  The ISM models of Redfield and Linsky predict the line-of-sight column density through the local ISM for an arbitrary look direction. }

Stars within 30 pc encompass almost all habitable zone planets that are candidates for spectroscopic characterization and biomarker detection (LUVOIR Team Final Report). ~\nocite{luvoir19}  \escape~ has the sensitivity to meet the required signal-to-noise and temporal resolution to address the key science questions (\S2) for F, G, K and active M stars out to 30 pc and inactive stars earlier than M5 out to ~8 pc. Thus, the interstellar medium is not the limiting factor in our ability to study EUV emission from the most important exoplanet hosts.

\escape~ obtains accurate measurements of the 80 – 560 \AA\ spectrum of nearby cool stars in the primary EUV channels.  The 570 – 911 \AA\ flux is also important for exoplanet heating calculations as this band overlaps with the peak of the ionization cross-sections of atomic and molecular hydrogen.  However, attenuation by the local ISM in the 570 – 911 \AA\ region prevents observations from all but the very nearest stars with the lowest ISM column density sightlines.   

\escape~ takes a two-tiered approach to measuring these important intermediate EUV wavelengths.  1) \escape~’s simultaneous coverage of coronal and transition region lines from 80 – 560 \AA\ and transition region and chromospheric lines in the FUV channel from 1300 – 1650 Å (e.g., C II, Si IV, C IV) provide robust constraints on differential emission measure calculations (DEM; see, e.g., Drake et al. 2020; Duvvuri et al. 2021 and references therein), allowing us to calculate accurate stellar fluxes in the 570 – 911 \AA\ region. 2) \escape~ includes a low-resolution mode covering 600 – 825 \AA\ with sufficient sensitivity to directly measure stellar emission in this bandpass, including the important H I Lyman continuum, for $\sim$10 stars in the DEEP survey.\nocite{drake20,duvvuri21}\footnote{ The 565 – 600 \AA\ and 825 – 911 \AA\ regions are avoided due to bright He I 584 \AA\ and O II 834 \AA\ geocoronal airglow emission,  respectively.}

\begin{figure}[htbp]
   \centering
   \includegraphics[scale=.4,clip,trim=0mm 0mm 5mm 5mm,angle=0]{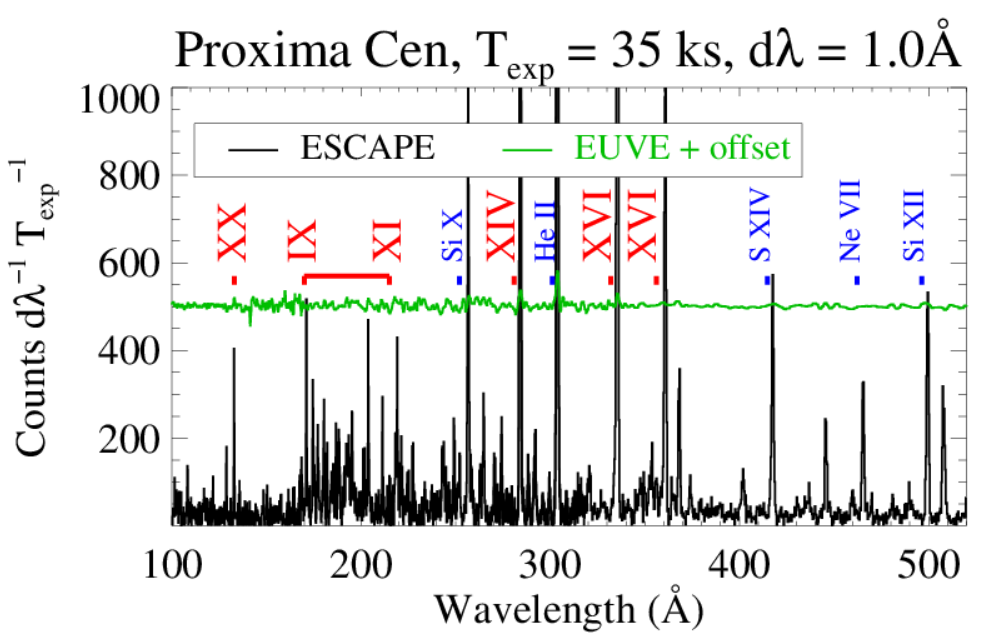}
   \caption{A simulated \escape~ SEEN spectrum of Proxima Cen is shown (black). The 77ks $EUVE$ exposure of Proxima Cen (green, offset by 500 counts) resulted in tentative detections of $\sim$5 stellar emission lines. \escape~ achieves a line-integrated S/N of $\sim$41 at Fe IX 171 \AA\ in 35 ksec. Prominent iron ionization states are labeled in red and other species in blue.   }
\end{figure}

\section{\escape~ Science Surveys: SEEN and DEEP}

\subsection{ Stellar Euv ENvironments (SEEN) Survey}
The Stellar Euv ENvironments (SEEN) survey measures the 80 – 1650 \AA\ irradiance and time variability for 200 stars with absolute photometric uncertainty $<$ 40\% in the  20 month baseline mission. \textcolor{black}{The photometric accuracy requirement is selected to remove the stellar flux as a dominant source of uncertainty in atmospheric escape calculations\cite{gronoff2020}.} We require the sensitivity and spectral resolution to separate stellar emission lines and measure variability in individual spectral features.  These requirements drive the telescope size and grazing incidence design to maintain high throughput at EUV wavelengths (Section 6).  To illustrate the comparison of a SEEN survey observation with archival data available from $EUVE$, we show a simulated raw spectrum of Proxima Cen in Figure 4.

  Approximately 20 flares are required to develop a flare-frequency distribution with $\sim$50\% logarithmic rate accuracy (e.g., the rate constant of the flare-frequency power-law distribution).  To determine the required SEEN survey duration that effectively constrains the flare rates of F, G, K, and M dwarfs, we estimate the EUV flare rate based on the FUV flare-frequency distribution of Loyd et al. (2018a) and the FUV-to-EUV scaling relations from \textcolor{black}{France et al. (2018)}~\cite{france18}.
  
  The flare frequency distribution is a power law, so more low energy flares are expected than high energy flares. Taking an intermediate activity M dwarf at 10 pc as our fiducial system, we find flares with total Fe IX 171 \AA\ energy $E$(Fe IX)~$\geq$~5~$\times$~10$^{29}$ erg produce a 5-$\sigma$ flare detection with \escape~ (assuming a typical 5-minute M dwarf UV flare duration).  This fiducial flare corresponds to a factor of $\sim$~15 brightening, typical for transition region flares observed on low-mass stars~\cite{hawley03,loyd18a,loyd18b}.
  
From the estimated EUV flare rate, we find that in a 35 ksec SEEN observation, we expect to detect 9 flares with 15$\times$ brightening $E$(Fe IX) = 10$^{29.75}$ ergs; 3 flares with 50$\times$ brightening $E$(Fe IX) = 10$^{30.25}$; and 1 flare with 100$\times$ brightening $E$(Fe IX) = 10$^{30.55}$.   To assemble statistically significant EUV flare-frequency distributions ($\geq$~20 flares),  we will group 3-5 stars with similar mass and age properties.  Flaring rates and fluxes of young G and K stars are comparable to intermediate activity M stars~\cite{ayres15,airapetian20}, therefore we expect young G and K stars to be well represented by the inactive M dwarf calculations presented above.

\begin{figure}[htbp]
   \centering
   \includegraphics[scale=.6,clip,trim=0mm 0mm 0mm 0mm,angle=0]{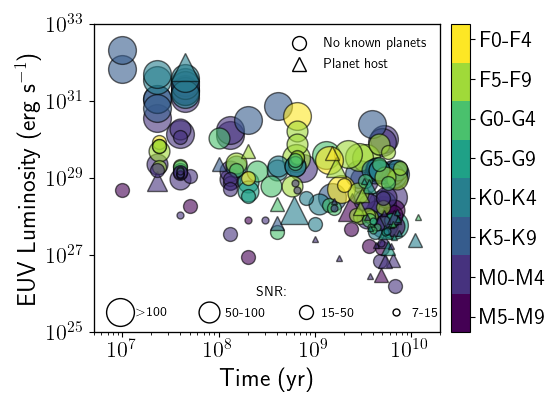}

   \caption{Estimated range of the EUV radiation field on stellar evolutionary timescales, characterized in \escape~'s SEEN survey. The EUV luminosity is the estimated 90 – 360 \AA\ intrinsic luminosity and the symbol sizes represent the integrated S/N of Fe IX in a 35 ksec SEEN observation.  Stars with confirmed exoplanets are shown as triangles. This plot shows our current 200 star target list, with resolution in both age and mass (color-coded).}

\end{figure}

\begin{figure*}
\centering
\includegraphics[width=1.0\textwidth]{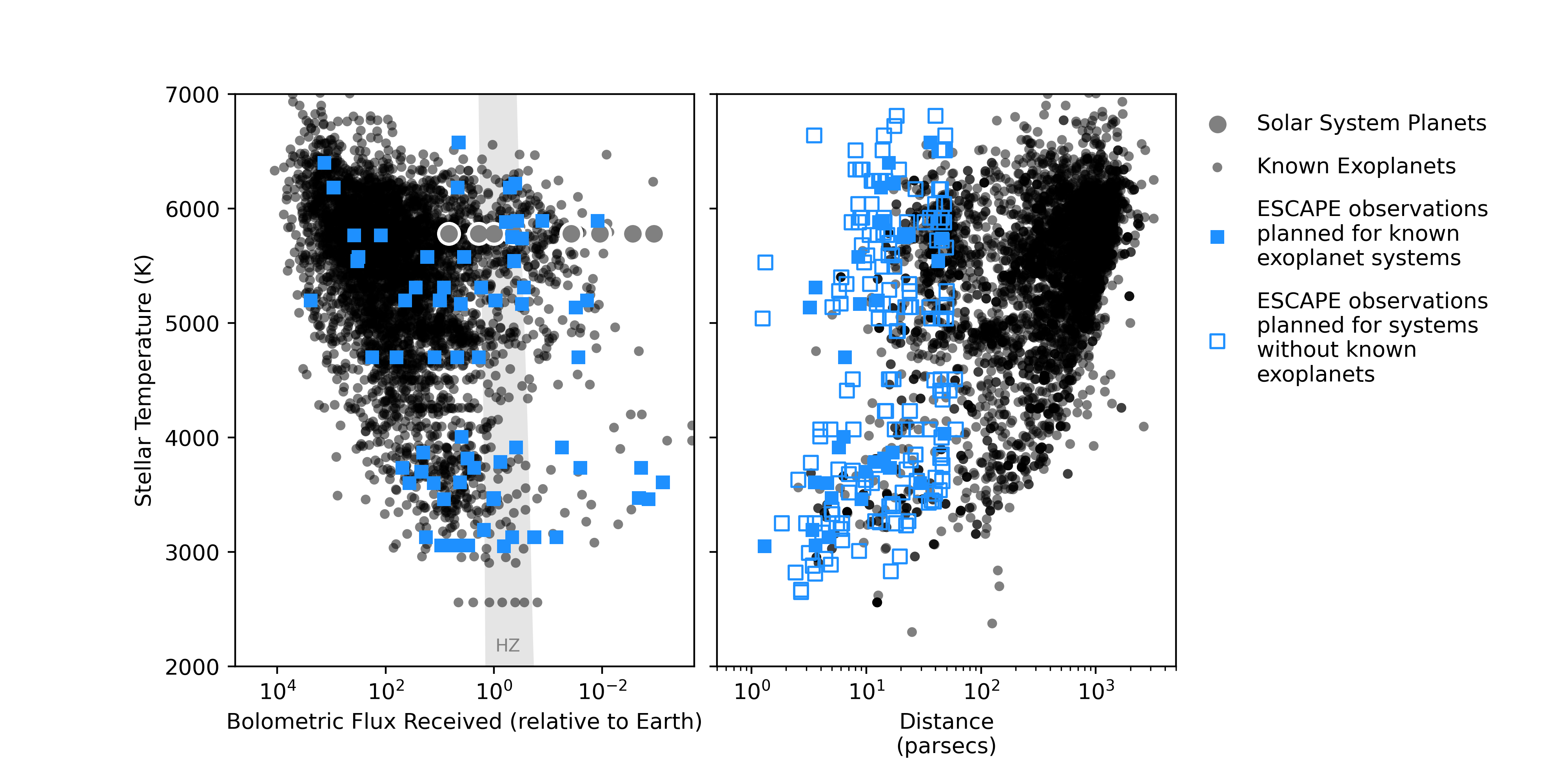}
    \caption{The \escape~ target list consists of 41 known exoplanet host stars, and is representative of the underlying known exoplanet population. {\it Left panel}: The known exoplanet population is shown as \textcolor{black}{bolometric flux received} against the host star’s effective temperature. {\it Right panel}: The known exoplanet population is shown as distance against stellar effective temperature. Solar system planets are shown as large gray/white circles, exoplanets are shown as small grey circles, and \escape~planned targets are blue squares (filled = known exoplanets, empty = no exoplanets yet known). An approximate estimate for the temperate zone (recent-Venus to early-Mars; Kopparapu et al. 2013) is shown as the gray shaded region.} 
    \label{OvsB}
\end{figure*}

\escape~ measures the critical pre-main sequence history of M dwarf EUV luminosities ~\cite{ribas17,pineda21}, as well as the evolution of the EUV luminosity of F, G, and K stars (e.g., Johnstone et al. 2021; Figure 5).  The SEEN Survey covers a range of ages of each spectral type, spanning pre-main-sequence stars in the TW Hya and $\beta$ Pic moving groups (ages $<$ 30 Myr), intermediate age clusters like the AB Dor and the Hyades (hundreds of Myr; Schneider et al. 2018), and `field' age stars comparable to the Sun (Gyr, based on gyrochronolgy and UV/optical activity indicators).  Each spectral type is broken down into approximate `early' (subclass 0 – 4) and `late' (subclass 5 – 9) subgroups, e.g., (early-F type stars (F0V – F4V), late-F type stars (F5V – F9V), early-G type stars (G0V – G4V), etc.).  The SEEN survey data allows us to assess how similar the EUV outputs of two stars of comparable mass and age are, and importantly, allows us to develop a range of EUV flux estimates for a given set of basic stellar parameters that provides empirical bounds for studies of exoplanets beyond the lifetime of the \escape~ mission.\nocite{kopparapu13,schneider18,johnstone21}

A full 200-star target list that meets these requirements at sufficient S/N is illustrated in Figures 5 and 6.  41 of the baseline stars host known planets (with 85 total planets, 15 of which are in the 200~--~330 K temperate zone).  While this target list satisfies \escape~’s science objectives, ongoing results from $TESS$ and $CHEOPS$ will inform the final target list that will be finalized prior to launch.  Finally, it should be noted that exoplanet populations statistics tell us that many of \escape~’s ``planetless" target stars likely host planets that are yet to be detected (Dressing \& Charbonneau 2015; Stark et al. 2019 and references therein).  Future direct-detection missions (e.g., LUVOIR and Habex) will both find and characterize these Earth-like planets around nearby solar-type (F, G, and K) stars.  \escape~’s SEEN survey ensures that when planets are directly imaged around these stars and spectroscopic characterization begins, the fundamental stellar constraints on the \textcolor{black}{long-term stability} of their atmospheres will be in place.~\nocite{dressing15,stark19}

\subsection{ Dedicated Euv Eruption Program (DEEP)}

The Dedicated Euv Eruption Program (DEEP) survey executes monitoring campaigns of 24 select stars to measure EUV flare frequency distributions and CME rates on individual, high-priority targets in the solar neighborhood.   


\escape~ measures the EUV flare frequency distribution of carefully selected F, G, K, and M dwarf systems on temporal baselines of $\sim$20–40 times longer than the best available datasets from $HST$.  Similarly, there has only been one comparable M dwarf campaign at X-ray wavelengths (0.15 – 15 keV; Kowalski e al. -- in prep), and these data do not sample the 10$^{5-6}$ K emission that contributes the majority of the stellar EUV flux~\cite{tilipman21}.  Archival $Kepler$ and $TESS$ observations of white-light flares may also provide connections between the optical and EUV frequencies of superflares (e.g., Loyd et al. 2018b; Howard et al. 2020).\nocite{loyd18b,howard20} 

The average CME rate of the Sun is $\sim$3.5 CME day$^{-1}$ and is $\sim$10 CME day$^{-1}$ at solar maximum \cite{gopalswamy09}. The Sun is less active than other stars of its type (e.g., Reinhold et al. 2020), so we conservatively estimate a CME rate of 5 CME day$^{-1}$ for Sun-like stars. Stellar CME rates are likely well-approximated by a Poisson distribution like the Sun, so the uncertainty on the CME frequency per day is the square root of the number of CMEs observed in a monitoring campaign divided by the monitoring duration.
Therefore, observing 9 CMEs results in a 3-$\sigma$ constraint on the CME rate. 
While not every CME is preceded by a dimming event, most are. Veronig et al. (2021) find that 97\% of CMEs have a corresponding irradiance dimming. However, only $\sim$~50\% of those will occur on the observable side of the star.\nocite{veronig21} 
Finally, approximately one-third of all solar dimming events meet \escape~’s 3-$\sigma$ spectrophotometric detection threshold (dimming depths $\geq$ 6\% in Fe IX 171~\AA\ and Fe X 174~\AA).
Combining the above numbers with \escape~’s DEEP survey
observing efficiency (78\%), the expected detection rate for solar-like CMEs with \escape~ is approximately 0.63 CMEs/day. \escape~ will monitor each star in the DEEP survey for $\approx$~14 days to provide sufficient temporal baselines to observe 9 CMEs/target, resulting in a 3-$\sigma$ determination of the CME rate per star, from which CME-driven atmospheric escape will be estimated.

\escape~'s spectral coverage and spectral resolution ($\Delta\lambda$~$<$~1.5 \AA) are designed to spectrally isolate specific emission lines for temporally-resolved lightcurve analysis over a range of coronal ionization states (Fe VIII – Fe XX). \escape~ has the sensitivity to detect solar-like dimming events (6\% max dimming depth in Fe IX 171 \AA\ and Fe X 174 \AA, summed over ten 30 minute exposures) to a distance of 6 pc. \escape~ will survey 10 F, G, and K stars for solar-like CMEs as part of the DEEP survey.  For these 10 F, G, and K stars, we expect \escape~ will detect over 90 solar-like coronal dimming events. Therefore, \escape~ will likely characterize the frequency of stellar CMEs for the first time.  Deeper dimming events or those detectable in broadband spectral binning~\cite{veronig21} are even more readily detectable in \escape~ time-series observations.  We note that while dimming events from the hot coronae associated with very active stars may be detectable in the X-rays, EUV spectral coverage is required to sample the lower quiescent coronal emission temperatures typical of the more numerous, lower activity, planet hosting stars.

Magnetic fluxes on active M stars can be factors of 10 – 1000 times higher than global solar magnetic fluxes~\cite{donati06,shulyak17},
potentially trapping the charged particle explosions associated with CMEs~\cite{alvarado18}.  Simulations by our team indicate that CMEs with energy $>$ 10$^{34}$ ergs will escape magnetic confinement from active M dwarfs and be detectable with \escape~ (Jin et al.~--~in prep).  \escape~’s observational capabilities enable us to search for EUV dimming signatures from M dwarfs for the first time, providing empirical constraints on the particle deposition into the atmospheres of orbiting planets.

\begin{figure*}[htpb]
\centering
\includegraphics[width=1.0\textwidth]{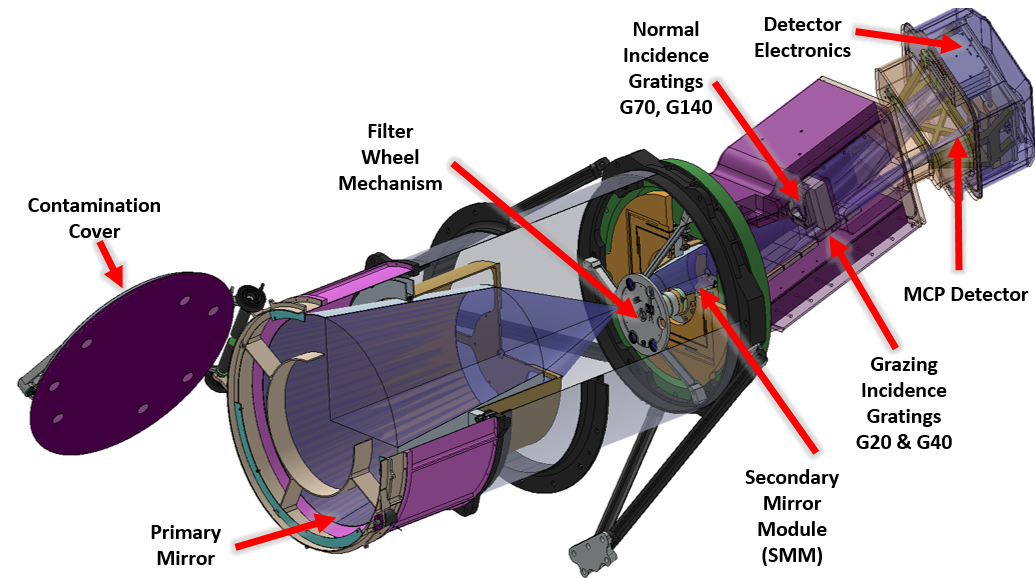}
    \caption{A schematic representation of the \escape~ science instrument: a grazing incidence Gregorian telescope (\S6.1.1) feeding a grazing incidence spectrograph (\S6.1.2). \textcolor{black}{The primary mirror diameter is 0.46-meters and the end-to-end length of the instrument is approximately 2-meters. } Major components are labeled.} 
    \label{OvsB}
\end{figure*}

\section{Implementation: The \escape~ Science Payload}

The \escape~ science mission requires measurements of the EUV and FUV fluxes of a sample of low-mass stars. To adequately carry out this program, \escape~ must be at least 25-times more efficient across the 100~--~500~\AA\ EUV band than any previous astrophysics spectrograph to operate in this range (EUVE DS/S; Bowyer \& Malina, 1991, Chandra LETG/HRC-S; Brinkman et al., 2000). \escape~’s efficiency gain compared to  previous missions is achieved with an instrument specifically designed to meet the \escape~ objectives; i.e., optimized to be a cool star characterization instrument.\nocite{bowyer91,brinkman00}


	The \escape~ telescope is the grazing incidence analog of a Gregorian (known as a “Hettrick-Bowyer;” HB, Hettrick \& Bowyer, 1984), with a 46 cm parabolic primary mirror, a prime focus aperture (PFA), and a re-focusing elliptical secondary. An accessible prime focus is a defining feature of Gregorian telescopes, and one of the critical elements that enables the high efficiency of \escape~; the field-limiting PFA provides stray-light suppression without costly bandpass limiting filters or additional reflections that would be prohibitive to meeting the system throughput requirements. The telescope is provided by a joint NASA/Marshall Space Flight Center (MSFC) and Smithsonian Astrophysical Observatory (SAO) team.   Figure 7 illustrates the \escape~ telescope and science instrument with key components labeled.\nocite{hettrick84}
	
	The \escape~ spectrograph intercepts the converging beam from the telescope with an array of etched silicon gratings supplied by Pennsylvania State University (PSU). The grating bank is divided into two channels to optimize performance over a broad EUV bandpass: the G20 (78 – 339 \AA) and the G40 (94 – 564 \AA; Table 1). The gratings are ruled with a quasi-radial groove pattern to maintain a constant linear  dispersion. Each channel is divided into segments sized to be compatible with a standard 6-inch silicon wafer for ease of manufacture and heritage.  Segments are co-aligned with the aid of an optical grating ruled onto a nominally unilluminated portion of the substrate.

	The resulting spectra are curved in an ``arc-of-diffraction” common in all radial groove spectrographs and recorded onto a microchannel plate (MCP) detector developed by the University of California, Berkeley (UCB) - the workhorse detector type for all EUV and FUV instruments. 	The zero-order reflections off of the G20 and G40 gratings are folded back onto two normal incidence re-focusing gratings, the G70 (600 – 825 \AA) and G140 (1250 – 1650 \AA), to obtain long-wave EUV and FUV coverage. These spectra are recorded onto the same MCP detector as the grazing channels, providing this spectral coverage with minimal additional complexity. The \escape~ instrument covers the required broad spectral range simultaneously with a fixed optical configuration that requires no mechanisms or bandpass limiting optics during normal science operations.

   \begin{table}[ht]
\caption{ESCAPE Spectroscopic Modes. } 
\label{tab:Multimedia-Specifications}
\begin{center}       
\begin{tabular}{c c c c c}
\hline
\rule[-1ex]{0pt}{3.5ex}  {\bf Grating} & {\bf Bandpass} & {\bf Spatial Resolution} & {\bf Spectral Resolution} & {\bf Peak A$_{eff}$}   \\
\rule[-1ex]{0pt}{3.5ex}  &  (\AA) & ($\mu$m) & (\AA) & (cm$^{2}$)   \\
\hline
\rule[-1ex]{0pt}{3.5ex}  G20 & 78~--~339 & 500 & 0.91$^{a}$ & 67.4 \\
\rule[-1ex]{0pt}{3.5ex}  G40 & 94~--~564 & 500 & 0.65$^{a}$ & 35.2  \\
\rule[-1ex]{0pt}{3.5ex}  G70 & 600~--~825 & 475  & 3.6$^{b}$ & 2.6  \\
\rule[-1ex]{0pt}{3.5ex}  G140 & 1250~--~1650 & 475  & 6.1$^{b}$ & 4.5  \\
\hline 
\end{tabular}
\end{center}
\begin{center}
$^{a}${Evaluated at 171~\AA.}\\
$^{b}${Average resolution over the band.}
\end{center}
\end{table}

\subsection{Instrument Subsystems}
In the following subsections, we describe the key components of the \escape~ instrument in greater detail. 

\subsubsection{\escape~ Telescope}

The \escape~ telescope is a grazing incidence Gregorian with a prime focus point in between the primary and secondary mirrors (see Figure 7; Hettrick \& Bowyer, 1984). The final architecture was chosen after a series of trade studies to maximize efficiency and minimize complexity. 

The \escape~ parabolic primary mirror is a 1 mm thick $\times$ 500 mm long, 46 cm maximum diameter nickel shell with a focus point at the prime focus aperture (PFA) pinhole (500 $\mu$m diameter). The PFA restricts the field-of-view of \escape~ to a top-hat function with a 1.6 arcminute full-transmission zone and wings extending out to a 3.6 arcminute diameter. This restricted field-of-view limits geocoronal airglow in the system without requiring costly bandpass limiting transmission filters. The diverging beam is re-focused by a 1 mm thick $\times$ 152 mm long, 13 cm outer diameter elliptical secondary mirror.

The mirror shells will be fashioned at NASA/MSFC by electro-forming a nickel-cobalt alloy onto a super-polished aluminum mandrel figured to the desired optic prescription. This is known as electro-formed nickel replication (ENR), the same mirror technology used for IXPE, $XMM-Newton$, FOXSI, MaGIXS, and other NASA flight programs. Both telescope optics are coated with 20 nm of zirconium in the coating facilities at SAO (Romaine et al., 2011). Zirconium has excellent performance at critical \escape~ wavelengths and is widely used for thin film applications. 



	
\subsubsection{\escape~ Spectrograph}
$Grazing~Incidence~Gratings$~--~
The \escape~ short-wavelength EUV channels utilize quasi-radial off-plane blazed gratings at grazing incidence~\cite{cash83,mcentaffer19}. Each channel intercepts roughly half of the telescope power at average graze angles of $\theta$= 11 degrees  (G20) and 19 degrees  (G40). The groove patterns are angled to match the converging telescope rays in order to maintain a constant linear dispersion, increasing the groove density as the rays approach focus. Light is diffracted along an “arc of diffraction,” with spectra offset in the cross-dispersion direction. The bandpass of each channel is extended by leveraging multiple spectral orders.  Higher spectral order contributions are suppressed by the telescope graze angles;  all m $\geq$ 3 order in G20 and m $\geq$ 4 in G40 have effective area (A$_{eff}$) $<$ 0.5 cm$^{2}$.

	The gratings are lithographically ruled into single crystal silicon wafers following a process developed by PSU under a NASA Strategic Astrophysics Technology (SAT) award~\cite{miles18} and demonstrated with \escape~ grating prescriptions~\cite{grise21}.  The wafers are cut from the parent boule so the $<$111$>$ crystal plane orientation matches the desired blaze angles. This grating fabrication technique produces smooth and uniform blaze facets, maximizing diffraction efficiency while readily meeting the low scatter requirement of $I$/$I_{o}$ $<$ 10$^{-3}$ at 10 \AA\ from an emission line to suppress scattered H I Ly$\alpha$. The segments are diced to a rectangular form and bonded to kovar mounts with $X$, $Y$, $\theta$ adjustment capability to aid in segment alignment.

$Normal~Incidence~Gratings$~--~
The NI gratings are supplied by Horiba Jobin-Yvon; the normal incidence channels leverage unused detector area to provide essential science capability with minimal added complexity.  The G70 and G140 channels consist of 14 mm square flat fold mirrors that pickoff the reflected zero-order light from the G40 and G20, respectively, and direct it to concave re-focusing holographic gratings. The spectra are imaged onto the same detector as the grazing incidence channels. The G140 channel is coated in MgF$_{2}$ protected aluminum, while the G70 channel is coated in silicon carbide (SiC). The gratings are bonded to an aluminum support structure with manual tip/tilt adjustment for alignment. 
The fold mirrors are mounted on the detector support truss in a small enclosure to baffle stray light. A 4 mm thick, 6 mm diameter CaF$_{2}$ window intercepts the G140 beam to attenuate $>$ 95\% of geocoronal H I Ly$\alpha$. 

\subsubsection{\escape~ Detector System}

The \escape~ spectra are imaged by a 125 $\times$ 40 mm MCP detector system with mild focal plane curvature (0.5m) and a cross delay line (XDL) readout anode. The baseline design uses borosilicate-glass MCPs activated by atomic layer deposition (ALD), which have flight heritage on large formats and with curvatures (DEUCE; Erickson et al., 2021, SISTINE; Cruz-Aguirre et al., 2021; JUNO-UVS; Davis et al., 2018).\nocite{erickson21,fleming16,france16b,davis18} 

 UV light sensing is accomplished with an opaque KBr photocathode; a standard option for EUV and FUV missions (including $EUVE$ and $FUSE$; Vallerga et al., 1994, Sahnow et al., 2000). Consistent with previous flight detectors, we employ a charged 95\% transmission grid to enhance the efficiency by collecting events that impinge on the detector face. The structure of this grid has \textcolor{black}{been examined as part of the $ESCAPE$ ray trace analysis and has been observed} to have no noticeable impact on the PSF. The XDL readout anode is based on the SISTINE and DEUCE multilayer designs (France et al. 2016; Erickson et al. 2021).\nocite{vallerga94,sahnow00} 

\escape~ utilizes an electronics package with flight heritage on numerous missions, including ICON-EUV, JUNO-UVS, JUICE-UVS, EUROPA-UVS and EMM-EMUS~\cite{darling15,davis11,davis19,retherford15}.
This package utilizes low-power amplifiers / discriminators for the start and stop signals of each axis of the XDL, followed by time to digital conversion, and an FPGA (ACTEL AX/RT family) to convert the raw event time of arrival differences into $x$, $y$ and signal amplitudes outputted as 32-bit LVDS (13-bit X, 12-bit Y, 7-bit pulse height). \textcolor{black}{Maximum global count rates for primary science targets, including airglow contributions, are less than 1000 Hz, posing no counting rate challenges for the detector system.}

\subsection{Instrument Performance}

The system effective  area is a function of the reflection efficiency of the optics ($R$), efficiency of the gratings ($\epsilon_g$), and quantum efficiency of the detector (DQE), multiplied by the geometric collecting area of the telescope (A$_{geo}$):
\begin{equation}
    A_{eff} (\lambda)=A_{geo}  R(\lambda,\theta) \epsilon_g (\lambda) DQE (\lambda)
\end{equation}

The reflectivity of each optic is a function of the wavelength and graze angle, $\theta$, which varies across the surface. To calculate this function, a grid of rays is traced through the system, with the graze angle at each optic intercept point used to calculate the power reflected as a function of wavelength. The rays are evenly spaced at the aperture to properly power-weight this function. 

\begin{figure}[htbp]
   \centering
   \includegraphics[scale=.5,clip,trim=0mm 0mm 5mm 5mm,angle=0]{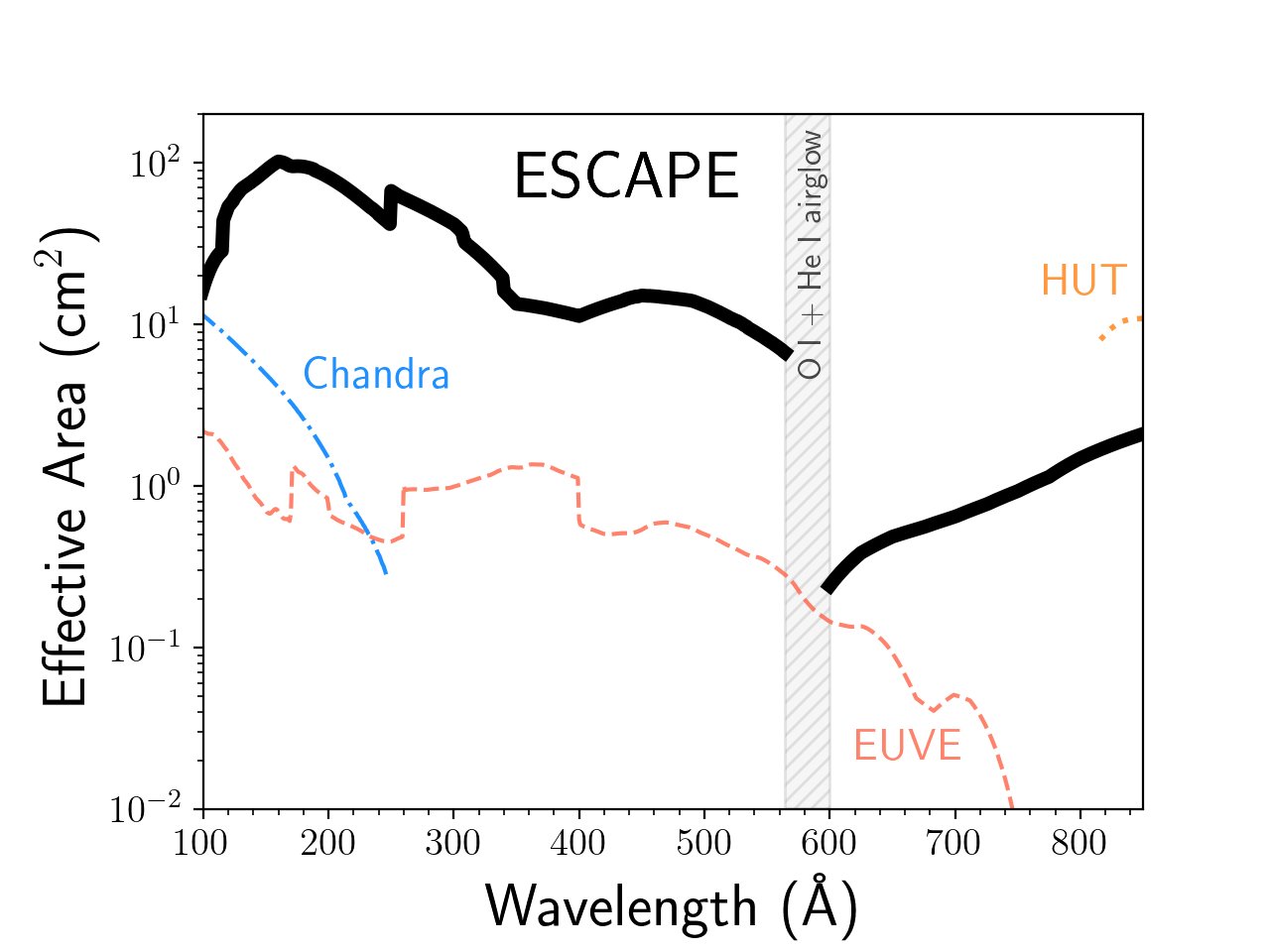}

   \caption{A comparison of the \escape~ effective area (solid black line) compared with previous astrophysics missions covering the 100~--~912~\AA\ EUV bandpass. \escape~ intentionally introduces a gap in spectral coverage between  approximately 550~--~600~\AA\ to avoid strong O II and He I geocoronal emission.}

\end{figure}

	The total A$_{eff}$ of \escape~ (Figure 8) is the sum of all diffraction orders, as well as the sum of the two GI channels for where the wavelength range overlaps. The A$_{eff}$ of the NI channels (G70, G140) is calculated with added terms for the fold mirror, CaF$_{2}$ filter (G140), and zero order reflection off of the corresponding GI grating. 

\begin{figure}[htbp]
   \centering
   \includegraphics[scale=.72,clip,trim=0mm 0mm 0mm 0mm,angle=0]{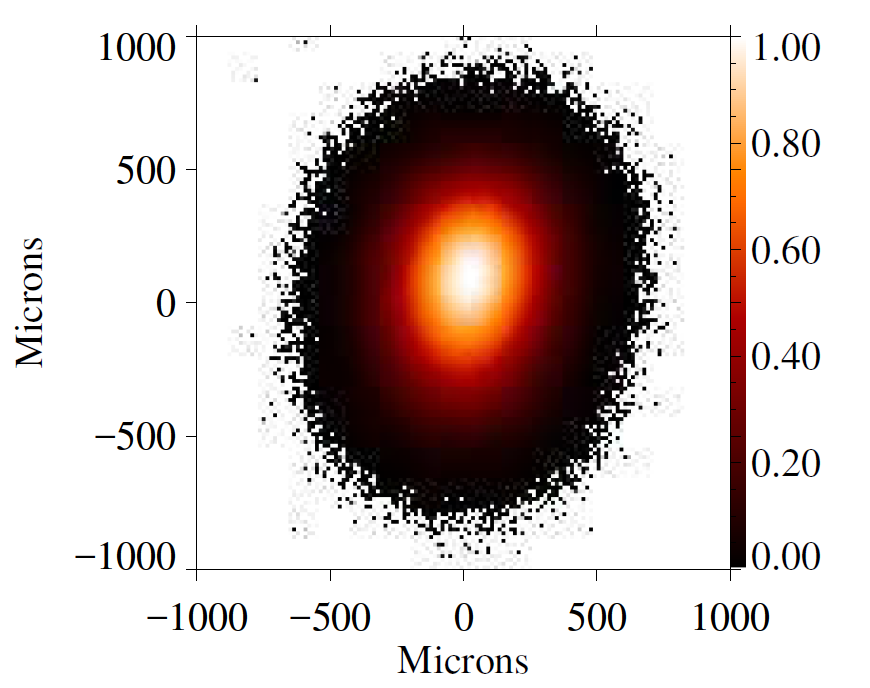}

   \caption{Projected PSF for the G20 mode at 171 \AA, sampled at the detector digital pixel scale \textcolor{black}{(approximately 14$\mu$m~$\times$~14$\mu$m)}.}

\end{figure}

The point-spread-function (PSF) is the sum of contributions from the intrinsic grating distortion, optical mount stresses, 1-sigma pointing jitter, maximum allowable mirror fabrication errors and maximum allowable detector resolution.  This PSF produces a predicted RMS spectral resolution performance at 171~\AA\ for the grazing incidence channels of 0.91~\AA\ in G20 and 0.66~\AA\ in G40 (second order), and normal incidence performance of 3.6~\AA\ in G70 and 6.1~\AA\ in G140 (see Table 1), averaged over the normal incidence mode bandpasses.
The raytraced PSF spot for the G20 at 171 \AA\ is presented in Figure 9.

\subsection{Simulated Data Example}

We carried out science performance calculations using a conservative spot size corresponding to the baseline  requirement of 1.5 \AA\ spectral resolution at 171~\AA, lower than the projected instrument performance.  A simulated two-dimensional detector image of a SEEN survey target is shown in Figure 10, $left$.  
\textcolor{black}{$ESCAPE$’s targets are cool stars, dominated by coronal and chromospheric line emission.  Therefore, the order overlap associated with continuum sources that was an issue for $EUVE$ is mitigated for $ESCAPE$ and wavelength identification can be accomplished without wavelength blocking filters.  For $ESCAPE$, the wavelength definition can be cleanly inferred from comparison of one-dimensional spectral extractions with solar or synthetic stellar spectra (as illustrated by the minimal spectral order overlap for solar-type stars shown in Figure 10, center panels). More sophisticated forward modeling techniques, used on grating spectra at X-ray wavelengths (see, e.g., Brinkman et al. 2000)\cite{brinkman00}, will be developed to identify spectral features by their location in the 2D spectrogram.  }

Spectral data are order-sorted by three methods: 1) laboratory spectra obtained during instrument integration and testing provide a map of spectral detector locations as a function of input wavelength for known calibration gases, 2) on-orbit checkout calibration observations of cool stars with $EUVE$ observations will be used to validate forward modeling analyses of coronal models; pre-flight simulations show clear delineation of different wavelengths in one-dimensional spectral traces for G20 and G40 channels (Figure 10, $center$), and 3) order-sorting filters (manufactured by Luxel) isolating short-, medium-, and long-wavelength EUV bands are well-established in solar missions and will be used to validate the forward modeling order sorting analysis as well as reject soft X-ray continuum leaking into the \escape~  band for stars with high-temperature coronae.  

An Al+Mg filter isolates the first order G40 spectrum (and supports the calculation of a predictive DEM model to guide the extraction).  The G40 ‘Open – Al+Mg’ difference spectra are cross-correlated with the Zr+MoSi$_{2}$ G20 spectra to isolate the first and second order G20 and G40 spectra.  An indium filter tests for short-wavelength bremsstrahlung leaking into the ESCAPE band.  Filters are only used for initial wavelength/order calibration activities; an open aperture position is used for all SEEN and DEEP survey observations.  

\begin{figure*}[t]
\centering
\includegraphics[width=0.95\textwidth]{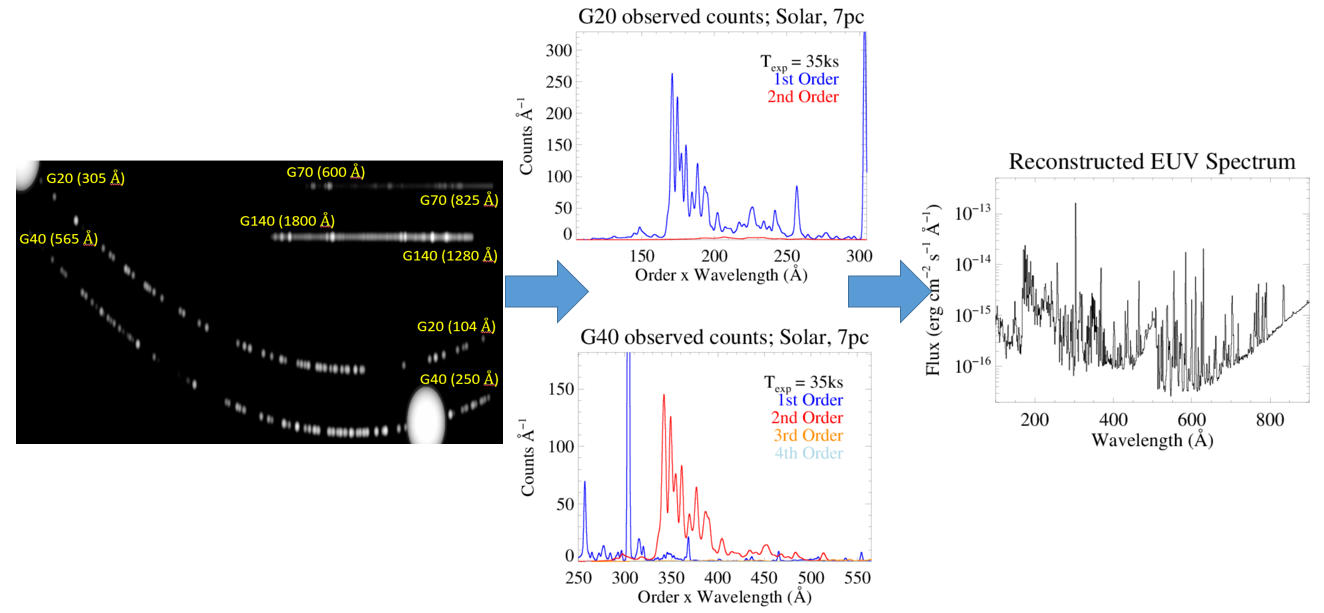}
    \caption{A schematic representation of the \escape~ science data flow for an inactive solar type star at a distance of 7 pc, with a typical SEEN exposure time of 35 ksec. ($left$) Level 3a science products contain the [$\lambda$, $y$, $t$] photon lists, including order identifications. \textcolor{black}{The oval features are He II 304~\AA\ airglow, which are spectrally isolated on the detector and do not impact the majority of the science spectrum.}  ($center$)  One-dimensional spectra are extracted and different wavelengths are readily identified in the G20 and G40 channels. ($right$) Order-sorted one-dimensional spectra are combined with interstellar attenuation corrections and differential emission measure models to produce full extreme ultraviolet irradiance spectra. } 
    \label{OvsB}
\end{figure*}

Figure 10 shows the typical data flow for \escape~ observations.  The Level 2 products contain the instrument data organized by observations, photon lists at full detector resolution, with all of the detector related corrections applied (application of ground-based flat-field, thermal and geometric distortion maps, and 2-D walk correction as necessary; e.g., COS Data Handbook 2015; France et al. 2011). The photon-list data are time-tagged to 100 msec accuracy.\nocite{france11} 

\begin{figure*}[htbp]
   \centering
    \includegraphics[scale=.6,clip,trim=0mm 0mm 5mm 5mm,angle=0]{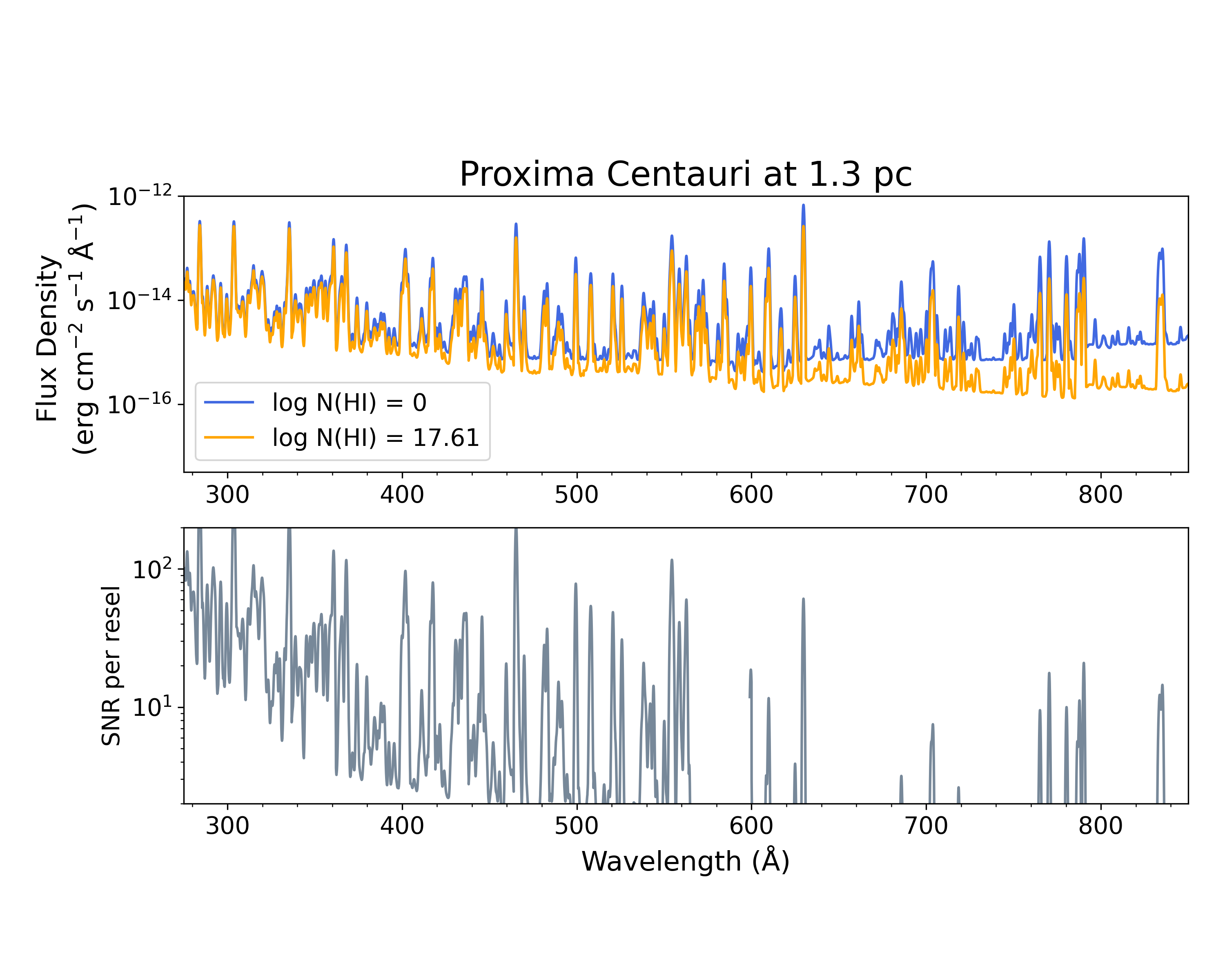}
   \caption{$top$~--~Simulated observation of the full EUV spectrum of Proxima Cen comparing the intrinsic (blue) and ISM-absorbed \escape~~data (orange).  $bottom$~--~Signal-to-noise per 1.5~\AA\  resolution element of a DEEP Survey observation of Proxima Cen.  }
\end{figure*}

Two Level 3 (L3) data products are generated: The L3a products preserve the full time-tag photon list from Level 2, with the addition of corresponding grating order probability {\textcolor{black}{(the fraction contributed by each order to the best-fit forward model spectrum at a given detector position})} and wavelength for the four expected orders. Wavelength solutions are provided to the end-user for the creation of lightcurves in individual spectral features (as is done for $HST$-COS).  Level 3a science products contain the [$\lambda$, $y$, $t$] photon lists that are required for stellar flare and CME characterization (Figure 10 $left$).  L3b contain the fully calibrated, one dimensional spectrum in ergs cm$^{-2}$ s$^{-1}$ \AA$^{-1}$ as a function of wavelength (in \AA) for each observation (both in individual orders and a final coadded spectrum). L3b data are combined with interstellar attenuation corrections and a DEM model to produce full extreme ultraviolet irradiance spectra (Figure 10, $right$).     The \escape~ data will be hosted at the MAST archive.

Level 3a and 3b data products will be produced for individual exposures and exposure time integrated for each \escape~ target to facilitate dissemination and use by the astronomical community.   Level 3b science products contain the full one-dimensional EUV and FUV stellar spectra that are required to measure and reconstruct the EUV irradiance incident on orbiting planets.  Figure 4 shows a simulated SEEN observation (extracted counts spectrum) of the nearby planet-hosting M dwarf Proxima Cen, compared to an archival $EUVE$ spectrum of the same source.  Figure 11 shows a Level 3b data product of a simulated DEEP observation (the flux spectrum) of Proxima Cen and the predicted signal-to-noise ratio as a function of wavelength that  includes the influence of local ISM absorption on the \escape~ observations.


\section{\escape~~Development Timeline}
\label{sec:conc}

ESCAPE completed \textcolor{black}{an initial} Phase A study in mid-2021, \textcolor{black}{but was not selected for full implementation.  \escape~ will be re-proposed to the next SMEX opportunity, presumably in $\sim$~2024.  If \escape~ is selected in that round, Phase B would begin in late 2025, culminating in the instrument and mission level preliminary design reviews in late 2026, respectively.   Phase C includes comprehensive design review in late Q2 2027 and extends through August 2028.   Phase D would run through the August 2029 launch and on-orbit commissioning.  On-orbit spectral resolution tests are made by observing the unresolved and spectrally isolated Fe XIV and Fe XVI lines as these spectral diagnostics are expected to be present in all cool star observations (coronal line widths are $<$ 10\% of instrumental resolution). Effective area as a function of wavelength is calibrated in flight by performing observations of well-studied hot white dwarf calibration stars.  Phase E comprises the 20 month science mission when the full SEEN and DEEP surveys are executed; ESCAPE maintains a 26\% mission lifetime margin to complete the baseline science mission. }

{\bf Acknowledgments:} The \escape~ Phase A mission was managed at the Laboratory for  Atmospheric and Space Physics at the University of Colorado Boulder under NASA contract 80GSFC21C0003.  Portions of this manuscript appeared as SPIE proceeding 11821-001.  The \escape~ team acknowledges the numerous invaluable discussions with colleagues on the need for and approach to an EUV spectroscopy mission over the previous five years.  In particular, KF thanks Alex Brown, Parke Loyd, and Tom Woods for enjoyable conversations on EUV spectroscopy and variability.   

\clearpage

\bibliography{escape_sci_bibtex_jatis} 

\begin{thebibliography}{100}

\bibitem{johnstone15}
C.~P. {Johnstone}, M.~{G{\"u}del}, A.~{St{\"o}kl}, {\em et~al.}, ``{The
  Evolution of Stellar Rotation and the Hydrogen Atmospheres of Habitable-zone
  Terrestrial Planets},'' {\em ApJ l} {\bf 815}, L12  (2015).

\bibitem{johnstone21}
C.~P. {Johnstone}, M.~{Bartel}, and M.~{G{\"u}del}, ``{The active lives of
  stars: A complete description of the rotation and XUV evolution of F, G, K,
  and M dwarfs},'' {\em A and A} {\bf 649}, A96  (2021).

\bibitem{tu15}
L.~{Tu}, C.~P. {Johnstone}, M.~{G{\"u}del}, {\em et~al.}, ``{The extreme
  ultraviolet and X-ray Sun in Time: High-energy evolutionary tracks of a
  solar-like star},'' {\em A and A} {\bf 577}, L3  (2015).

\bibitem{amerstorfer17}
U.~V. {Amerstorfer}, H.~{Gr{\"o}ller}, H.~{Lichtenegger}, {\em et~al.},
  ``{Escape and evolution of Mars's CO$_{2}$ atmosphere: Influence of
  suprathermal atoms},'' {\em Journal of Geophysical Research (Planets)} {\bf
  122}, 1321--1337  (2017).

\bibitem{vidal04}
A.~{Vidal-Madjar}, J.-M. {D{\'e}sert}, A.~{Lecavelier des Etangs}, {\em
  et~al.}, ``{Detection of Oxygen and Carbon in the Hydrodynamically Escaping
  Atmosphere of the Extrasolar Planet HD 209458b},'' {\em ApJ l} {\bf 604},
  L69--L72  (2004).

\bibitem{linsky10}
J.~L. {Linsky}, H.~{Yang}, K.~{France}, {\em et~al.}, ``{Observations of Mass
  Loss from the Transiting Exoplanet HD 209458b},'' {\em ApJ} {\bf 717},
  1291--1299  (2010).

\bibitem{ballester15}
G.~E. {Ballester} and L.~{Ben-Jaffel}, ``{Re-visit of HST FUV Observations of
  the Hot-Jupiter System HD 209458: No Si III Detection and the Need for COS
  Transit Observations},'' {\em ApJ} {\bf 804}, 116  (2015).

\bibitem{fossati10}
L.~{Fossati}, C.~A. {Haswell}, C.~S. {Froning}, {\em et~al.}, ``{Metals in the
  Exosphere of the Highly Irradiated Planet WASP-12b},'' {\em ApJ l} {\bf 714},
  L222--L227  (2010).

\bibitem{haswell12}
C.~A. {Haswell}, L.~{Fosatti}, and A.~{et al.}, ``{Near-UV Absorption,
  Chromospheric Activity, and Star-Planet Interactions in the WASP-12
  system},'' {\em ApJ} {\bf 0}, 1--2  (2012).

\bibitem{sing19}
D.~K. {Sing}, P.~{Lavvas}, G.~E. {Ballester}, {\em et~al.}, ``{The Hubble Space
  Telescope PanCET Program: Exospheric Mg II and Fe II in the Near-ultraviolet
  Transmission Spectrum of WASP-121b Using Jitter Decorrelation},'' {\em AJ}
  {\bf 158}, 91  (2019).

\bibitem{cubillos20}
P.~E. {Cubillos}, L.~{Fossati}, T.~{Koskinen}, {\em et~al.},
  ``{Near-ultraviolet Transmission Spectroscopy of HD 209458b: Evidence of
  Ionized Iron Beyond the Planetary Roche Lobe},'' {\em AJ} {\bf 159}, 111
  (2020).

\bibitem{kulikov06}
Y.~N. {Kulikov}, H.~{Lammer}, H.~I.~M. {Lichtenegger}, {\em et~al.},
  ``{Atmospheric and water loss from early Venus},'' {\em PLANSS} {\bf 54},
  1425--1444  (2006).

\bibitem{lichtnegger16}
H.~I.~M. {Lichtenegger}, K.~G. {Kislyakova}, P.~{Odert}, {\em et~al.}, ``{Solar
  XUV and ENA-driven water loss from early Venus' steam atmosphere},'' {\em
  Journal of Geophysical Research (Space Physics)} {\bf 121}, 4718--4732
  (2016).

\bibitem{dong17}
C.~{Dong}, M.~{Lingam}, Y.~{Ma}, {\em et~al.}, ``{Is Proxima Centauri b
  Habitable? A Study of Atmospheric Loss},'' {\em ApJ l} {\bf 837}, L26
  (2017).

\bibitem{airapetian20}
V.~S. {Airapetian}, R.~{Barnes}, O.~{Cohen}, {\em et~al.}, ``{Impact of space
  weather on climate and habitability of terrestrial-type exoplanets},'' {\em
  International Journal of Astrobiology} {\bf 19}, 136--194  (2020).

\bibitem{gronoff2020}
G.~Gronoff, P.~Arras, S.~Baraka, {\em et~al.}, ``Atmospheric escape processes
  and planetary atmospheric evolution,'' {\em Journal of Geophysical Research:
  Space Physics} {\bf 125}(8), e2019JA027639  (2020).
\newblock e2019JA027639 10.1029/2019JA027639.

\bibitem{ramirez14}
R.~M. {Ramirez} and L.~{Kaltenegger}, ``{The Habitable Zones of
  Pre-main-sequence Stars},'' {\em ApJ l} {\bf 797}, L25  (2014).

\bibitem{luger15}
R.~{Luger} and R.~{Barnes}, ``{Extreme Water Loss and Abiotic O2Buildup on
  Planets Throughout the Habitable Zones of M Dwarfs},'' {\em Astrobiology}
  {\bf 15}, 119--143  (2015).

\bibitem{wordsworth18}
R.~D. {Wordsworth}, L.~K. {Schaefer}, and R.~A. {Fischer}, ``{Redox Evolution
  via Gravitational Differentiation on Low-mass Planets: Implications for
  Abiotic Oxygen, Water Loss, and Habitability},'' {\em AJ} {\bf 155}, 195
  (2018).

\bibitem{france16}
K.~{France}, R.~O.~P. {Loyd}, A.~{Youngblood}, {\em et~al.}, ``{The MUSCLES
  Treasury Survey. I. Motivation and Overview},'' {\em ApJ} {\bf 820}, 89
  (2016).

\bibitem{drake20}
J.~J. {Drake}, V.~L. {Kashyap}, B.~J. {Wargelin}, {\em et~al.}, ``{Pointing
  Chandra toward the Extreme Ultraviolet Fluxes of Very Low Mass Stars},'' {\em
  ApJ} {\bf 893}, 137  (2020).

\bibitem{linsky14}
J.~L. {Linsky}, J.~{Fontenla}, and K.~{France}, ``{The Intrinsic Extreme
  Ultraviolet Fluxes of F5 V TO M5 V Stars},'' {\em ApJ} {\bf 780}, 61  (2014).

\bibitem{forcada11}
J.~{Sanz-Forcada}, G.~{Micela}, I.~{Ribas}, {\em et~al.}, ``{Estimation of the
  XUV radiation onto close planets and their evaporation},'' {\em A and A} {\bf
  532}, A6  (2011).

\bibitem{ribas17}
I.~{Ribas}, E.~{Bolmont}, F.~{Selsis}, {\em et~al.}, ``{The habitability of
  Proxima Centauri b. I. Irradiation, rotation and volatile inventory from
  formation to the present},'' {\em A and A} {\bf 596}, A111  (2016).

\bibitem{loyd16}
R.~O.~P. {Loyd}, K.~{France}, A.~{Youngblood}, {\em et~al.}, ``{The MUSCLES
  Treasury Survey. III. X-Ray to Infrared Spectra of 11 M and K Stars Hosting
  Planets},'' {\em ApJ} {\bf 824}, 102  (2016).

\bibitem{woods09}
T.~N. {Woods}, P.~C. {Chamberlin}, J.~W. {Harder}, {\em et~al.}, ``{Solar
  Irradiance Reference Spectra (SIRS) for the 2008 Whole Heliosphere Interval
  (WHI)},'' {\em GRL} {\bf 36}, 1101  (2009).

\bibitem{fontenla16}
J.~M. {Fontenla}, J.~L. {Linsky}, J.~{Witbrod}, {\em et~al.}, ``{Semi-empirical
  Modeling of the Photosphere, Chromosphere, Transition Region, and Corona of
  the M-dwarf Host Star GJ 832},'' {\em ApJ} {\bf 830}, 154  (2016).

\bibitem{garcia17}
K.~{Garcia-Sage}, A.~{Glocer}, J.~J. {Drake}, {\em et~al.}, ``{On the Magnetic
  Protection of the Atmosphere of Proxima Centauri b},'' {\em ApJ l} {\bf 844},
  L13  (2017).

\bibitem{airapetian17}
V.~S. {Airapetian}, A.~{Glocer}, G.~V. {Khazanov}, {\em et~al.}, ``{How
  Hospitable Are Space Weather Affected Habitable Zones? The Role of Ion
  Escape},'' {\em ApJ l} {\bf 836}, L3  (2017).

\bibitem{bowyer91}
S.~{Bowyer} and R.~F. {Malina}, ``{The extreme ultraviolet explorer mission},''
  {\em Advances in Space Research} {\bf 11}, 205--215  (1991).

\bibitem{drake95}
J.~J. {Drake}, J.~M. {Laming}, and K.~G. {Widing}, ``{Stellar Coronal
  Abundances. II. The First Ionization Potential Effect and Its Absence in the
  Corona of Procyon},'' {\em ApJ} {\bf 443}, 393  (1995).

\bibitem{drake97}
J.~J. {Drake}, J.~M. {Laming}, and K.~G. {Widing}, ``{Stellar Coronal
  Abundances. V. Evidence for the First Ionization Potential Effect in
  {\ensuremath{\alpha}} Centauri},'' {\em ApJ} {\bf 478}, 403--416  (1997).

\bibitem{walkowicz09}
L.~M. {Walkowicz} and S.~L. {Hawley}, ``{Tracers of Chromospheric Structure. I.
  Observations of Ca II K and H{$\alpha$} in M Dwarfs},'' {\em AJ} {\bf 137},
  3297--3313  (2009).

\bibitem{woods12}
T.~N. {Woods}, F.~G. {Eparvier}, R.~{Hock}, {\em et~al.}, ``{Extreme
  Ultraviolet Variability Experiment (EVE) on the Solar Dynamics Observatory
  (SDO): Overview of Science Objectives, Instrument Design, Data Products, and
  Model Developments},'' {\em Solar Physics} {\bf 275}, 115--143  (2012).

\bibitem{segura10}
A.~{Segura}, L.~M. {Walkowicz}, V.~{Meadows}, {\em et~al.}, ``{The Effect of a
  Strong Stellar Flare on the Atmospheric Chemistry of an Earth-like Planet
  Orbiting an M Dwarf},'' {\em Astrobiology} {\bf 10}, 751--771  (2010).

\bibitem{davenport16}
J.~R.~A. {Davenport}, ``{The Kepler Catalog of Stellar Flares},'' {\em ApJ}
  {\bf 829}, 23  (2016).

\bibitem{guenther20}
M.~N. {G{\"u}nther}, Z.~{Zhan}, S.~{Seager}, {\em et~al.}, ``{Stellar Flares
  from the First TESS Data Release: Exploring a New Sample of M Dwarfs},'' {\em
  AJ} {\bf 159}, 60  (2020).

\bibitem{howard20}
W.~S. {Howard}, H.~{Corbett}, N.~M. {Law}, {\em et~al.}, ``{EvryFlare. III.
  Temperature Evolution and Habitability Impacts of Dozens of Superflares
  Observed Simultaneously by Evryscope and TESS},'' {\em ApJ} {\bf 902}, 115
  (2020).

\bibitem{hawley91}
S.~L. {Hawley} and B.~R. {Pettersen}, ``{The great flare of 1985 April 12 on AD
  Leonis},'' {\em ApJ} {\bf 378}, 725--741  (1991).

\bibitem{venot16}
O.~{Venot}, M.~{Rocchetto}, S.~{Carl}, {\em et~al.}, ``{Influence of Stellar
  Flares on the Chemical Composition of Exoplanets and Spectra},'' {\em ApJ}
  {\bf 830}, 77  (2016).

\bibitem{audard00}
M.~{Audard}, M.~{G{\"u}del}, J.~J. {Drake}, {\em et~al.},
  ``{Extreme-Ultraviolet Flare Activity in Late-Type Stars},'' {\em ApJ} {\bf
  541}, 396--409  (2000).

\bibitem{jones10}
B.~W. {Jones} and P.~N. {Sleep}, ``{Habitability of exoplanetary systems with
  planets observed in transit},'' {\em MNRAS} {\bf 407}, 1259--1267  (2010).

\bibitem{france20}
K.~{France}, G.~{Duvvuri}, H.~{Egan}, {\em et~al.}, ``{The High-energy
  Radiation Environment around a 10 Gyr M Dwarf: Habitable at Last?},'' {\em
  AJ} {\bf 160}, 237  (2020).

\bibitem{hawley03}
S.~L. {Hawley}, J.~C. {Allred}, C.~M. {Johns-Krull}, {\em et~al.},
  ``{Multiwavelength Observations of Flares on AD Leonis},'' {\em ApJ} {\bf
  597}, 535--554  (2003).

\bibitem{loyd14}
R.~O.~P. {Loyd} and K.~{France}, ``{Fluctuations and Flares in the Ultraviolet
  Line Emission of Cool Stars: Implications for Exoplanet Transit
  Observations},'' {\em ApJ s} {\bf 211}, 9  (2014).

\bibitem{loyd18a}
R.~O.~P. {Loyd}, K.~{France}, A.~{Youngblood}, {\em et~al.}, ``{The MUSCLES
  Treasury Survey. V. FUV Flares on Active and Inactive M Dwarfs},'' {\em ApJ}
  {\bf 867}, 71  (2018).

\bibitem{west15}
A.~A. {West}, K.~L. {Weisenburger}, J.~{Irwin}, {\em et~al.}, ``{An
  Activity-Rotation Relationship and Kinematic Analysis of Nearby
  Mid-to-Late-Type M Dwarfs},'' {\em ApJ} {\bf 812}, 3  (2015).

\bibitem{khodachenko07}
M.~L. {Khodachenko}, I.~{Ribas}, H.~{Lammer}, {\em et~al.}, ``{Coronal Mass
  Ejection (CME) Activity of Low Mass M Stars as An Important Factor for The
  Habitability of Terrestrial Exoplanets. I. CME Impact on Expected
  Magnetospheres of Earth-Like Exoplanets in Close-In Habitable Zones},'' {\em
  Astrobiology} {\bf 7}, 167--184  (2007).

\bibitem{lammer07}
H.~{Lammer}, H.~I.~M. {Lichtenegger}, Y.~N. {Kulikov}, {\em et~al.}, ``{Coronal
  Mass Ejection (CME) Activity of Low Mass M Stars as An Important Factor for
  The Habitability of Terrestrial Exoplanets. II. CME-Induced Ion Pick Up of
  Earth-like Exoplanets in Close-In Habitable Zones},'' {\em Astrobiology} {\bf
  7}, 185--207  (2007).

\bibitem{cherenkov17}
A.~{Cherenkov}, D.~{Bisikalo}, L.~{Fossati}, {\em et~al.}, ``{The Influence of
  Coronal Mass Ejections on the Mass-loss Rates of Hot-Jupiters},'' {\em ApJ}
  {\bf 846}, 31  (2017).

\bibitem{mason14}
J.~P. {Mason}, T.~N. {Woods}, A.~{Caspi}, {\em et~al.}, ``{Mechanisms and
  Observations of Coronal Dimming for the 2010 August 7 Event},'' {\em ApJ}
  {\bf 789}, 61  (2014).

\bibitem{mason16}
J.~P. {Mason}, T.~N. {Woods}, D.~F. {Webb}, {\em et~al.}, ``{Relationship of
  EUV Irradiance Coronal Dimming Slope and Depth to Coronal Mass Ejection Speed
  and Mass},'' {\em ApJ} {\bf 830}, 20  (2016).

\bibitem{mason19}
J.~P. {Mason}, R.~{Attie}, C.~N. {Arge}, {\em et~al.}, ``{The SDO/EVE Solar
  Irradiance Coronal Dimming Index Catalog. I. Methods and Algorithms},'' {\em
  ApJ s} {\bf 244}, 13  (2019).

\bibitem{tilley19}
M.~A. {Tilley}, A.~{Segura}, V.~{Meadows}, {\em et~al.}, ``{Modeling Repeated M
  Dwarf Flaring at an Earth-like Planet in the Habitable Zone: Atmospheric
  Effects for an Unmagnetized Planet},'' {\em Astrobiology} {\bf 19}, 64--86
  (2019).

\bibitem{chassefiere96}
E.~{Chassefi{\`e}re}, ``{Hydrodynamic Escape of Oxygen from Primitive
  Atmospheres: Applications to the Cases of Venus and Mars},'' {\em Icarus}
  {\bf 124}, 537--552  (1996).

\bibitem{jakosky15}
B.~M. {Jakosky}, J.~M. {Grebowsky}, J.~G. {Luhmann}, {\em et~al.}, ``{MAVEN
  observations of the response of Mars to an interplanetary coronal mass
  ejection},'' {\em Science} {\bf 350}, 0210  (2015).

\bibitem{jakosky18}
B.~M. {Jakosky}, D.~{Brain}, M.~{Chaffin}, {\em et~al.}, ``{Loss of the Martian
  atmosphere to space: Present-day loss rates determined from MAVEN
  observations and integrated loss through time},'' {\em Icarus} {\bf 315},
  146--157  (2018).

\bibitem{yoshiro06}
S.~{Yashiro}, S.~{Akiyama}, N.~{Gopalswamy}, {\em et~al.}, ``{Different
  Power-Law Indices in the Frequency Distributions of Flares with and without
  Coronal Mass Ejections},'' {\em ApJ l} {\bf 650}, L143--L146  (2006).

\bibitem{osten15}
R.~A. {Osten} and S.~J. {Wolk}, ``{Connecting Flares and Transient Mass-loss
  Events in Magnetically Active Stars},'' {\em ApJ} {\bf 809}, 79  (2015).

\bibitem{odert17}
P.~{Odert}, M.~{Leitzinger}, A.~{Hanslmeier}, {\em et~al.}, ``{Stellar coronal
  mass ejections - I. Estimating occurrence frequencies and mass-loss rates},''
  {\em MNRAS} {\bf 472}, 876--890  (2017).

\bibitem{crosley18}
M.~K. {Crosley} and R.~A. {Osten}, ``{Constraining Stellar Coronal Mass
  Ejections through Multi-wavelength Analysis of the Active M Dwarf EQ Peg},''
  {\em ApJ} {\bf 856}, 39  (2018).

\bibitem{haisch83}
B.~M. {Haisch}, J.~L. {Linsky}, P.~L. {Bornmann}, {\em et~al.}, ``{Coordinated
  Einstein and IUE observations of a disparitions brusques type flare event and
  quiescent emission from Proxima Centauri.},'' {\em ApJ} {\bf 267}, 280--290
  (1983).

\bibitem{moschou17}
S.-P. {Moschou}, J.~J. {Drake}, O.~{Cohen}, {\em et~al.}, ``{A Monster CME
  Obscuring a Demon Star Flare},'' {\em ApJ} {\bf 850}, 191  (2017).

\bibitem{harra16}
L.~K. {Harra}, C.~J. {Schrijver}, M.~{Janvier}, {\em et~al.}, ``{The
  Characteristics of Solar X-Class Flares and CMEs: A Paradigm for Stellar
  Superflares and Eruptions?},'' {\em Solar Physics} {\bf 291}, 1761--1782
  (2016).

\bibitem{woods11}
T.~N. {Woods}, R.~{Hock}, F.~{Eparvier}, {\em et~al.}, ``{New Solar
  Extreme-ultraviolet Irradiance Observations during Flares},'' {\em ApJ} {\bf
  739}, 59  (2011).

\bibitem{veronig21}
A.~M. {Veronig}, P.~{Odert}, M.~{Leitzinger}, {\em et~al.}, ``{Indications of
  stellar coronal mass ejections through coronal dimmings},'' {\em Nature
  Astronomy}   (2021).

\bibitem{sterling97}
A.~C. {Sterling} and H.~S. {Hudson}, ``{Yohkoh SXT Observations of X-Ray
  ``Dimming'' Associated with a Halo Coronal Mass Ejection},'' {\em ApJ l} {\bf
  491}, L55--L58  (1997).

\bibitem{dissauer18a}
K.~{Dissauer}, A.~M. {Veronig}, M.~{Temmer}, {\em et~al.}, ``{On the Detection
  of Coronal Dimmings and the Extraction of Their Characteristic Properties},''
  {\em ApJ} {\bf 855}, 137  (2018).

\bibitem{dissauer18b}
K.~{Dissauer}, A.~M. {Veronig}, M.~{Temmer}, {\em et~al.}, ``{Statistics of
  Coronal Dimmings Associated with Coronal Mass Ejections. I. Characteristic
  Dimming Properties and Flare Association},'' {\em ApJ} {\bf 863}, 169
  (2018).

\bibitem{aschwanden09}
M.~J. {Aschwanden}, N.~V. {Nitta}, J.-P. {Wuelser}, {\em et~al.}, ``{First
  Measurements of the Mass of Coronal Mass Ejections from the EUV Dimming
  Observed with STEREO EUVI A+B Spacecraft},'' {\em ApJ} {\bf 706}, 376--392
  (2009).

\bibitem{reinard09}
A.~A. {Reinard} and D.~A. {Biesecker}, ``{The Relationship between Coronal
  Dimming and Coronal Mass Ejection Properties},'' {\em ApJ} {\bf 705},
  914--919  (2009).

\bibitem{redfield08}
S.~{Redfield} and J.~L. {Linsky}, ``{The Structure of the Local Interstellar
  Medium. IV. Dynamics, Morphology, Physical Properties, and Implications of
  Cloud-Cloud Interactions},'' {\em ApJ} {\bf 673}, 283--314  (2008).

\bibitem{lehner03}
N.~{Lehner}, E.~B. {Jenkins}, C.~{Gry}, {\em et~al.}, ``{Far Ultraviolet
  Spectroscopic Explorer Survey of the Local Interstellar Medium within 200
  Parsecs},'' {\em ApJ} {\bf 595}, 858--879  (2003).

\bibitem{wood05}
B.~E. {Wood}, S.~{Redfield}, J.~L. {Linsky}, {\em et~al.}, ``{Stellar
  Ly{$\alpha$} Emission Lines in the Hubble Space Telescope Archive: Intrinsic
  Line Fluxes and Absorption from the Heliosphere and Astrospheres},'' {\em ApJ
  s} {\bf 159}, 118--140  (2005).

\bibitem{youngblood16}
A.~{Youngblood}, K.~{France}, R.~O.~P. {Loyd}, {\em et~al.}, ``{The MUSCLES
  Treasury Survey. II. Intrinsic LY{\ensuremath{\alpha}} and Extreme
  Ultraviolet Spectra of K and M Dwarfs with Exoplanets*},'' {\em ApJ} {\bf
  824}, 101  (2016).

\bibitem{france16b}
K.~{France}, K.~{Hoadley}, B.~T. {Fleming}, {\em et~al.}, ``{The SLICE, CHESS,
  and SISTINE Ultraviolet Spectrographs: Rocket-Borne Instrumentation
  Supporting Future Astrophysics Missions},'' {\em Journal of Astronomical
  Instrumentation} {\bf 5}, 1640001  (2016).

\bibitem{wood21}
B.~E. {Wood}, H.-R. {M{\"u}ller}, S.~{Redfield}, {\em et~al.}, ``{New
  Observational Constraints on the Winds of M dwarf Stars},'' {\em Apj} {\bf
  915}, 37  (2021).

\bibitem{luvoir19}
{LUVOIR Final Report}, ``{LUVOIR: Telling the Story of Life in the Universe},''
  {\em NASA}   (2019).

\bibitem{duvvuri21}
G.~M. {Duvvuri}, J.~{Sebastian Pineda}, Z.~K. {Berta-Thompson}, {\em et~al.},
  ``{Reconstructing the Extreme Ultraviolet Emission of Cool Dwarfs Using
  Differential Emission Measure Polynomials},'' {\em ApJ} {\bf 913}, 40
  (2021).

\bibitem{france18}
K.~{France}, N.~{Arulanantham}, L.~{Fossati}, {\em et~al.}, ``{Far-ultraviolet
  Activity Levels of F, G, K, and M Dwarf Exoplanet Host Stars},'' {\em ApJ s}
  {\bf 239}, 16  (2018).

\bibitem{loyd18b}
R.~O.~P. {Loyd}, E.~L. {Shkolnik}, A.~C. {Schneider}, {\em et~al.}, ``{HAZMAT.
  IV. Flares and Superflares on Young M Stars in the Far Ultraviolet},'' {\em
  ApJ} {\bf 867}, 70  (2018).

\bibitem{ayres15}
T.~R. {Ayres}, ``{The Flare-ona of EK Draconis},'' {\em AJ} {\bf 150}, 7
  (2015).

\bibitem{pineda21}
J.~S. {Pineda}, A.~{Youngblood}, and K.~{France}, ``{The Far Ultraviolet
  M-dwarf Evolution Survey. I. The Rotational Evolution of High-energy
  Emissions},'' {\em ApJ} {\bf 911}, 111  (2021).

\bibitem{kopparapu13}
R.~K. {Kopparapu}, ``{A Revised Estimate of the Occurrence Rate of Terrestrial
  Planets in the Habitable Zones around Kepler M-dwarfs},'' {\em ApJ l} {\bf
  767}, L8  (2013).

\bibitem{schneider18}
A.~C. {Schneider} and E.~L. {Shkolnik}, ``{HAZMAT. III. The UV Evolution of
  Mid- to Late-M Stars with GALEX},'' {\em AJ} {\bf 155}, 122  (2018).

\bibitem{dressing15}
C.~D. {Dressing} and D.~{Charbonneau}, ``{The Occurrence of Potentially
  Habitable Planets Orbiting M Dwarfs Estimated from the Full Kepler Dataset
  and an Empirical Measurement of the Detection Sensitivity},'' {\em ApJ} {\bf
  807}, 45  (2015).

\bibitem{stark19}
C.~C. {Stark}, R.~{Belikov}, M.~R. {Bolcar}, {\em et~al.}, ``{ExoEarth yield
  landscape for future direct imaging space telescopes},'' {\em Journal of
  Astronomical Telescopes, Instruments, and Systems} {\bf 5}, 024009  (2019).

\bibitem{tilipman21}
D.~{Tilipman}, M.~{Vieytes}, J.~L. {Linsky}, {\em et~al.}, ``{Semiempirical
  Modeling of the Atmospheres of the M Dwarf Exoplanet Hosts GJ 832 and GJ
  581},'' {\em ApJ} {\bf 909}, 61  (2021).

\bibitem{gopalswamy09}
N.~{Gopalswamy}, S.~{Yashiro}, G.~{Michalek}, {\em et~al.}, ``{The SOHO/LASCO
  CME Catalog},'' {\em Earth Moon and Planets} {\bf 104}, 295--313  (2009).

\bibitem{donati06}
J.-F. {Donati}, T.~{Forveille}, A.~{Collier Cameron}, {\em et~al.}, ``{The
  Large-Scale Axisymmetric Magnetic Topology of a Very-Low-Mass Fully
  Convective Star},'' {\em Science} {\bf 311}, 633--635  (2006).

\bibitem{shulyak17}
D.~{Shulyak}, A.~{Reiners}, A.~{Engeln}, {\em et~al.}, ``{Strong dipole
  magnetic fields in fast rotating fully convective stars},'' {\em Nature
  Astronomy} {\bf 1}, 0184  (2017).

\bibitem{alvarado18}
J.~D. {Alvarado-G{\'o}mez}, J.~J. {Drake}, O.~{Cohen}, {\em et~al.},
  ``{Suppression of Coronal Mass Ejections in Active Stars by an Overlying
  Large-scale Magnetic Field: A Numerical Study},'' {\em ApJ} {\bf 862}, 93
  (2018).

\bibitem{brinkman00}
B.~C. {Brinkman}, T.~{Gunsing}, J.~S. {Kaastra}, {\em et~al.}, ``{Description
  and performance of the low-energy transmission grating spectrometer on board
  Chandra},'' in {\em X-Ray Optics, Instruments, and Missions III},  J.~E.
  {Truemper} and B.~{Aschenbach}, Eds., {\em Society of Photo-Optical
  Instrumentation Engineers (SPIE) Conference Series} {\bf 4012}, 81--90
  (2000).

\bibitem{hettrick84}
M.~C. {Hettrick} and S.~{Bowyer}, ``{Grazing incidence telescopes: a new class
  for soft X-ray and EUV spectroscopy.},'' {\em Applied Optics} {\bf 23},
  3732--3735  (1984).

\bibitem{cash83}
J.~{Cash}, Webster~C., ``{X-ray spectrographs using radial groove gratings},''
  {\em Applied Optics} {\bf 22}, 3971--3976  (1983).

\bibitem{mcentaffer19}
R.~L. {McEntaffer}, ``{Reflection Grating Spectrographs for Astrophysical
  Missions: Rockets, Explorers, Probes, and Flagship},'' in {\em The Space
  Astrophysics Landscape for the 2020s and Beyond},   {\bf 2135}, 5039  (2019).

\bibitem{miles18}
D.~M. {Miles}, J.~A. {McCoy}, R.~L. {McEntaffer}, {\em et~al.}, ``{Fabrication
  and Diffraction Efficiency of a Large-format, Replicated X-Ray Reflection
  Grating},'' {\em ApJ} {\bf 869}, 95  (2018).

\bibitem{grise21}
F.~{Gris{\'e}}, R.~{McEntaffer}, D.~{Miles}, {\em et~al.}, ``{Opening the road
  to custom astronomical UV gratings},'' in {\em American Astronomical Society
  Meeting Abstracts \#235},  {\em American Astronomical Society Meeting
  Abstracts} {\bf 235}, 373.16  (2020).

\bibitem{erickson21}
N.~{Erickson}, J.~{Green}, N.~{Nell}, {\em et~al.}, ``{DEUCE: a sounding-rocket
  ultraviolet spectrograph for flux-calibrated B star observations across the
  Lyman limit},'' {\em Journal of Astronomical Telescopes, Instruments, and
  Systems} {\bf 7}, 015002  (2021).

\bibitem{fleming16}
B.~T. {Fleming}, K.~{France}, N.~{Nell}, {\em et~al.}, ``{SISTINE: a pathfinder
  for FUV imaging spectroscopy on future NASA astrophysics missions},'' in {\em
  Space Telescopes and Instrumentation 2016: Ultraviolet to Gamma Ray},
  J.-W.~A. {den Herder}, T.~{Takahashi}, and M.~{Bautz}, Eds., {\em Society of
  Photo-Optical Instrumentation Engineers (SPIE) Conference Series} {\bf 9905},
  99050A  (2016).

\bibitem{davis18}
M.~W. {Davis}, G.~R. {Gladstone}, T.~K. {Greathouse}, {\em et~al.}, ``{Stray
  and scattered light properties of the Juno ultraviolet spectrograph},'' in
  {\em Space Telescopes and Instrumentation 2018: Ultraviolet to Gamma Ray},
  J.-W.~A. {den Herder}, S.~{Nikzad}, and K.~{Nakazawa}, Eds., {\em Society of
  Photo-Optical Instrumentation Engineers (SPIE) Conference Series} {\bf
  10699}, 106990J  (2018).

\bibitem{vallerga94}
J.~V. {Vallerga}, M.~{Eckert}, M.~{Sirk}, {\em et~al.}, ``{Long-term orbital
  performance of the microchannel plate (MCP) detectors aboard the Extreme
  Ultraviolet Explorer},'' in {\em EUV, X-Ray, and Gamma-Ray Instrumentation
  for Astronomy V},  O.~H. {Siegmund} and J.~V. {Vallerga}, Eds., {\em Society
  of Photo-Optical Instrumentation Engineers (SPIE) Conference Series} {\bf
  2280}, 57--68  (1994).

\bibitem{sahnow00}
D.~J. {Sahnow}, H.~W. {Moos}, T.~B. {Ake}, {\em et~al.}, ``{On-Orbit
  Performance of the Far Ultraviolet Spectroscopic Explorer Satellite},'' {\em
  ApJ l} {\bf 538}, L7--L11  (2000).

\bibitem{darling15}
N.~T. {Darling}, O.~H.~W. {Siegmund}, T.~{Curtis}, {\em et~al.}, ``{Performance
  results of the ICON FUV sealed tube converters},'' in {\em UV, X-Ray, and
  Gamma-Ray Space Instrumentation for Astronomy XIX},  O.~H. {Siegmund}, Ed.,
  {\em Society of Photo-Optical Instrumentation Engineers (SPIE) Conference
  Series} {\bf 9601}, 96010T  (2015).

\bibitem{davis11}
M.~W. {Davis}, G.~R. {Gladstone}, T.~K. {Greathouse}, {\em et~al.},
  ``{Radiometric performance results of the Juno ultraviolet spectrograph
  (Juno-UVS)},'' in {\em Society of Photo-Optical Instrumentation Engineers
  (SPIE) Conference Series},  {\em Society of Photo-Optical Instrumentation
  Engineers (SPIE) Conference Series} {\bf 8146}, 814604  (2011).

\bibitem{davis19}
M.~W. {Davis}, O.~H.~W. {Siegmund}, G.~R. {Gladstone}, {\em et~al.}, ``{Bench
  and thermal vacuum testing of the JUICE-UVS microchannel plate detector
  system},'' in {\em UV, X-Ray, and Gamma-Ray Space Instrumentation for
  Astronomy XXI},  {\em Society of Photo-Optical Instrumentation Engineers
  (SPIE) Conference Series} {\bf 11118}, 111180Q  (2019).

\bibitem{retherford15}
K.~D. {Retherford}, T.~K. {Greathouse}, G.~R. {Gladstone}, {\em et~al.}, ``{The
  Far-UV Albedo of the Moon as a Probe of the Lunar Cryosphere: LRO Lyman Alpha
  Mapping Project (LAMP) Latest Results},'' in {\em Lunar and Planetary Science
  Conference},  {\em Lunar and Planetary Science Conference}, 2213  (2015).

\bibitem{france11}
K.~{France}, R.~{McCray}, S.~V. {Penton}, {\em et~al.}, ``{HST-COS Observations
  of Hydrogen, Helium, Carbon, and Nitrogen Emission from the SN 1987A Reverse
  Shock},'' {\em ApJ} {\bf 743}, 186  (2011).

\end{thebibliography}
\bibliographystyle{spiejour}   


\vspace{2ex}\noindent\textbf{Kevin France} is a professor in the Department of Astrophysical and Planetary Sciences at the University of Colorado.  Dr. France’s research focuses on exoplanets and their host stars, protoplanetary disks, and the development of instrumentation for ultraviolet astrophysics.   He is the Principal Investigator of the ESCAPE Small Explorer mission, the CUTE small satellite mission, and a NASA-supported sounding rocket program to study stellar impacts on exoplanet atmospheres and flight-test critical path hardware for future UV/optical astrophysics missions.  He is a regular guest observer with the Hubble Space Telescope and the chair of the Space Telescope Users Committee.  He has been a member of the HST-COS instrument and science teams and the LUVOIR Science and Technology Definition Team.  Dr. France received his Ph.D. from the Johns Hopkins University in 2006 and was awarded NASA’s Nancy Grace Roman Fellowship in 2013.

\vspace{1ex}

\listoffigures
\listoftables

\end{spacing}
\end{document}